\documentclass[aps, pra, 10pt, twocolumn, groupedaddress, nofootinbib, longbibliography]{revtex4-1}

\usepackage[british]{babel}
\usepackage{graphicx}
\usepackage{times}

\usepackage{amssymb}
\usepackage{physics} 
\usepackage{mathtools}
\usepackage{siunitx}
\usepackage[dvipsnames]{xcolor}
\definecolor{mygreen}{RGB}{44,85,17}
\definecolor{myblue}{RGB}{34,31,150}
\definecolor{myred}{RGB}{255,66,56}

\usepackage[breaklinks=true,colorlinks=true,linkcolor=myblue,urlcolor=myblue,citecolor=myblue]{hyperref}

\makeatletter
\newcommand*{\transpose}{%
  {\mathpalette\@transpose{}}%
}
\newcommand*{\@transpose}[2]{%
  \raisebox{\depth}{$\m@th#1\intercal$}%
}
\makeatother

\begin{document}

% Title, authors, research centres and date

\title{Quantum scale estimation}
\author{Jes\'{u}s Rubio}
\email{J.Rubio-Jimenez@exeter.ac.uk}
\affiliation{Department of Physics and Astronomy, University of Exeter, Stocker Road, Exeter EX4 4QL, UK}

\date{\today}
   
\begin{abstract}

Quantum scale estimation, as introduced and explored here, establishes the most precise framework for the estimation of scale parameters that is allowed by the laws of quantum mechanics. 
This addresses an important gap in quantum metrology, since current practice focuses almost exclusively on the estimation of phase and location parameters.
For given prior probability and quantum state, and using Bayesian principles, a rule to construct the optimal probability-operator measurement is provided. 
Furthermore, the corresponding minimum mean logarithmic error is identified. 
This is then generalised as to accommodate the simultaneous estimation of multiple scale parameters, and a procedure to classify practical measurements into optimal, almost-optimal or sub-optimal is highlighted. 
As a means of illustration, the new framework is exploited to generalise scale-invariant global thermometry, as well as to address the estimation of the lifetime of an atomic state.
On a more conceptual note, the optimal strategy is employed to construct an observable for scale parameters, an approach which may serve as a template for a more systematic search of quantum observables.  
Quantum scale estimation thus opens a new line of enquire---the precise measurement of scale parameters such as temperatures and rates---within the quantum information sciences. 

\end{abstract}

 \maketitle

% Body of the document

%---------------------------------------------------------

\section{Introduction}

Quantum estimation theory is typically envisioned as a toolbox to design highly precise measurements \cite{helstrom1976, paris2009, rafal2015, demkowicz2020, sidhu2020}.
Discovered over half a century ago \cite{helstrom1967mmse, helstrom1968multiparameter, personick1969thesis, personick1971, helstrom1974, helstrom1976, yuen1973, holevo1973, holevo1973b}, this framework is currently attracting a great deal of attention due to its central role in `the second quantum revolution' \cite{dowling2003}---an application of the principles of quantum mechanics to sensing, metrology, communication, computation, nano and space technologies \cite{dunningham2006, vinjanampathy2016, barnett2017, browne2017, degen2017, acin2018, belenchia2021}. 
But, beyond its unquestionable value as a practical tool, estimation theory may be attributed a more prominent role within the foundations of physics: it enables the means to connect, in a clear and systematic fashion, the variables that we measure in the laboratory with the variables appearing in our theories of nature. Indeed, if $x$ represents one of the possible outcomes that a given measurement could generate, one can often postulate the existence of a map $x \mapsto \tilde{\theta}(x)$ from the outcome $x$ to an estimate $\tilde{\theta}(x)$ for a theoretical variable $\Theta$ \cite{helstrom1976,holevo2011,kay1993,jaynes2003}. 
To find such map in practice, an elegant and neat approach is to minimise a mean error $\bar{\epsilon} = \langle \mathcal{D}[\tilde{\theta}(x),\theta] \rangle $ over the estimator function $\tilde{\theta}(x)$, where $\theta$ is a hypothesis about the true value of $\Theta$, $\mathcal{D}$ gauges the deviation of $\tilde{\theta}(x)$ from $\theta$, and the brackets indicate a weighted average over $(x, \theta)$. 
An instance of $\mathcal{D}$ is the familiar square error $\mathcal{D}(\tilde{\theta},\theta) = 
(\tilde{\theta} - \theta)^2$ \cite{helstrom1976,jaynes2003}. 

Estimation-theoretic quantities can, in addition, transcend their purely statistical role to acquire a new function within physics.
For example, mean errors are known to enable the derivation of generalised uncertainty relations \cite{helstrom1976,braunstein1996, 1996brody,holevo2011,watanabe2011,mankei2011,downes2017} such as those for phase-number \cite{helstrom1976, braunstein1996, jesus2020mar}, time-energy \cite{braunstein1996,brody1997} and temperature-energy \cite{miller2018, rubio2020} variables, and they appear to lead to a more detailed account of the notion of compatibility in quantum mechanics \cite{sammy2016compatibility, carollo2019, albarelli2019, albarelli2019novB, mankei2020, demkowicz2020, sidhu2021,candeloro2021}. 
Moreover, estimators provide the means to construct quantum observables such as those for phase \cite{personick1971,shapiro1991} and time \cite{holevo2011} parameters. 
Perhaps more importantly, estimation theory offers the possibility of treating dynamical variables and parameters alike in an unified manner \cite{holevo2011, helstrom1976}; this is demonstrated by the estimation-theoretic study of position variables in Ref.~\cite[Ch.~8]{helstrom1976}. 

In light of this fundamental outlook, selecting a deviation function $\mathcal{D}(\tilde{\theta},\theta)$ cannot be purely a matter of choice or convenience, but should instead be informed by physical reasoning. 
This is apparent in phase estimation \cite{pezze2014, rafal2015}, where the periodic nature of optical phases imposes certain constraints on which functions $\mathcal{D}$ are consistent in that context \cite{holevo2011, rafal2015}. 
The sine error $\mathcal{D}(\tilde{\theta},\theta) = 4\,\sin^2[(\tilde{\theta} - \theta)/2 ]$, in particular, is often employed for this purpose \cite{helstrom1976, berry2000, chiara2003, demkowicz2011, gebhart2021}, and while there are other functions $\mathcal{D}$ that are appropriate for phase estimation \cite{berry2000, holevo2011}, it is instructive to appreciate that the square error $\mathcal{D}(\tilde{\theta},\theta) = 
(\tilde{\theta} - \theta)^2$ is generally not one of them.\footnote{
The square error can be used as an \emph{approximation} to the sine error when the estimator $\tilde{\theta}$ and the values for the hypothesis $\theta$ are close \cite{rafal2015, jesus2018}.} 

In thermometry \cite{prosper1993, stace2010quantum, jahnke2011, DePasquale2018, potts2019fundamental, mok2020, rubio2020, alves2021, mehboudi2021,jorgensen2021bayesian,boeyens2021noninformative}, the unknown parameter, temperature, can sometimes be seen as establishing the relevant energy scale \cite{prosper1993, gregory2005}. 
Such a scenario calls for an estimation technique that preserves scale invariance \cite{jaynes1968, jaynes2003, toussaint2011}, and this idea has be shown to lead to a logarithmic deviation function $\mathcal{D}(\tilde{\theta},\theta) = \log^2(\tilde{\theta}/ \theta)$ \cite{rubio2020}. The implications and scope of this tool within quantum thermometry are currently under analysis \cite{mehboudi2021,jorgensen2021bayesian,boeyens2021noninformative}, including the role of adaptive measurements \cite{mehboudi2021} and its connection with the notion of thermodynamic length \cite{jorgensen2021bayesian}. 
However,  the logarithmic error transcends its thermometric origin, in the sense that it is also valid for other quantities playing the role of a scale \cite{rubio2020}. 
For example, to estimate Poisson rates \cite{jaynes1968, jaynes2003}, kinetic parameters \cite{baaske2014,subramanian2020,subramanian2021,eerqing2021,mpofu2021,mpofu2021exp}, and decay rates \cite{2009barnett}.
We may then ask: how can quantum mechanics enable, on the basis of logarithmic errors, the most precise estimation of scale parameters?

The present work provides a definite answer by establishing a quantum framework for the estimation scale parameters.
Once Sec.~\ref{sec:formulation} formulates the problem, Sec.~\ref{sec:main_result} derives an analytical expression for the true minimum mean logarithmic error.
Not only does this recover the optimal estimator found in Ref.~\cite{rubio2020}, but, notably, it further provides a rule to calculate the associated probability-operator measurement (POM)\footnote{
Also known as positive operator-valued measure (POVM) \cite{nielsen2010}.
} that is optimal for a given quantum state.
Moreover, this framework can incorporate prior knowledge, as it is built on Bayesian principles \cite{jaynes2003, toussaint2011}.

To illustrate these results, Sec.~\ref{sec:example} considers two applications: thermometry, and the estimation of the lifetime of an atomic state.
Within the context of equilibrium thermometry \cite{DePasquale2018,mehboudi2019}, and assuming discrete but otherwise arbitrary spectra, the question of whether energy measurements are generally optimal---i.e., for an arbitrarily large prior temperature range---is answered in the affirmative. 
This is not only of independent interest,\footnote{
In particular, this endows the error bound for energy measurements on a gas of spin-$1/2$ particles in Ref.~\cite{rubio2020} with a fundamental character, provided that adaptive measurements are excluded \cite{mehboudi2021}.
} but it also shows that the abstract POM which is optimal for quantum scale estimation can correspond to a practical measurement strategy.

Sec.~\ref{sec:discussion} proceeds then to explore some implications of the new toolbox. 
Sec.~\ref{sec:Smeaning} and \ref{sec:obs} construct a quantum observable for scale parameters, and the procedure leading to it is highlighted as a systematic method to construct other observables. 
Sec.~\ref{sec:fund-bounds} rewrites the true minimum as an inequality statement to assess whether a given practical POM is optimal, almost-optimal or sub-optimal. 
Sec.~\ref{sec:multiparameter} applies the new framework to the simultaneous estimation of multiple scale parameters, identifying a lower bound on a multiparameter mean logarithmic error.
Sec.~\ref{sec:invariance} closes with a discussion of invariance arguments in estimation theory.

To conclude, Sec.~\ref{sec:conclusions} argues that quantum scale estimation, when taken together with phase estimation and those results based on the mean square error, completes a trio of estimation theories for three of the most elementary quantities that one could possibly measure: phases, locations and scales. 

\section{Formulation of the problem}\label{sec:formulation}

Let an experiment be described by the following quantities: a measurand $x$, an unknown parameter $\Theta$, and a collection of known parameters $\boldsymbol{y} = (y_1, y_2, \dots)$.
Construct the vector $\boldsymbol{z} = (x,\boldsymbol{y})$. 
We say that $\Theta$ scales $z_i$---the $i$-th component of $\boldsymbol{z}$---if, for fixed $\Theta$, $z_i$ is considered ‘large' when $z_i/\Theta \gg 1$ and ‘small' when $z_i/\Theta \ll 1$, provided that $\Theta$ and $z_i$ are positive and measured in the same units.
The key aspect of this definition is its invariance under transformations $\lbrace z_i \mapsto z_i' =\, \gamma z_i, \Theta \mapsto \Theta' =\, \gamma \Theta \rbrace$,
with positive $\gamma$, since $z_i/\Theta = z_i'/\Theta'$.
This is the notion of \emph{scale parameter}---referring to $\Theta$---employed in this work.
We shall then call \emph{scale estimation} to any procedure aimed at retrieving an unknown scale parameter. 

Now suppose we wish to estimate $\Theta$, under the following assumptions: (i) knowledge of $\boldsymbol{y}$ does not inform the plausibility of the possible values for $\Theta$; (ii) $\Theta$ scales all the parameters $\boldsymbol{y}$, but not the measurand $x$, which is dimensionless; and (iii) $\Theta$ is encoded within a quantum state.

To proceed, we first construct a prior probability $p(\theta|\boldsymbol{y})$ encoding what is known about $\Theta$ irrespective of the measurement process, where it is recalled that $\theta$ denotes a hypothesis about the true value of $\Theta$.
Assumption (i) allows us to take $p(\theta|\boldsymbol{y}) \mapsto p(\theta)$ \cite{jaynes2003}. 
The form of $p(\theta)$ is determined either by context-specific prior knowledge or, should this not be available, by imposing maximum ignorance on the basis of assumption (ii) \cite{jaynes2003, demkowicz2020}, which leads to Jeffreys's prior $p(\theta) \propto 1/\theta$ \cite{jeffreys1961, jaynes1968}.

The next step is to find a likelihood model $p(x|\theta, \boldsymbol{y})$ linking the measurand $x$ with the unknown parameter $\Theta$.
Assumption (ii) can be enforced by imposing $p(x |\theta,\boldsymbol{y}) = p(x|\theta',\boldsymbol{y}')$.
This leads to $ p(x |\theta,\boldsymbol{y}) = p(x|\gamma\theta,\gamma\boldsymbol{y})$, which is a functional equation with solution $p(x|\theta,\boldsymbol{y}) = h(x,\boldsymbol{y}/\theta)$ \cite[Ch.~3]{aczel1987}, for some function $h$. 
That is, the likelihood for a given measurement outcome is an arbitrary function of the ratios $(y_1/\Theta, y_2/\Theta, \dots)$.

At the same time, assumption (iii), together with the Born rule, implies $p(x|\theta, \boldsymbol{y}) = \mathrm{Tr}[M_{\boldsymbol{y}}(x) \rho_{\boldsymbol{y}}(\theta)]$, where $\rho_{\boldsymbol{y}}(\theta)$ is a density operator and $M_{\boldsymbol{y}}(x)$ is a POM. 
Hence, the likelihood function is 
\begin{align}
p(x|\theta, \boldsymbol{y}) = \mathrm{Tr}[M_{\boldsymbol{y}}(x) \rho_{\boldsymbol{y}}(\theta)] = h\hspace{-0.1em}\left(x,\frac{\boldsymbol{y}}{\theta}\right),
\label{eq:likelihood-construction}
\end{align}
with the specific shape of $h$ given by detailed knowledge of the experimental design. 
As we can see, Eq.~\eqref{eq:likelihood-construction} imposes a constraint on both the state and the POM, and it defines a broad class of models for which the framework of quantum scale estimation in this work applies. 
For a discussion of other scale estimation problems that can arise when assumptions (ii) and (iii) are relaxed, see Sec.~\ref{sec:other-laws}.

Equipped with the prior probability and the likelihood function above, we can invoke Bayes theorem \cite{jaynes2003,toussaint2011} to calculate the posterior probability 
\begin{equation}
    p(\theta|x,\boldsymbol{y}) \propto p(\theta) \mathrm{Tr}[M_{\boldsymbol{y}}(x) \rho_{\boldsymbol{y}}(\theta)] = p(\theta)\,h\hspace{-0.1em}\left(x,\frac{\boldsymbol{y}}{\theta}\right),
    \label{eq:posterior}
\end{equation}
which fully solves the problem of estimating $\Theta$.
Although we could stop here, it is often useful to summarise the key features of Eq.~\eqref{eq:posterior} via a point estimator $\tilde{\theta}_{\boldsymbol{y}}
(x)$ with measurement-dependent uncertainty\footnote{
For a discussion of this type of errors in quantum metrology, see Ref.~\cite{hanamura2021}.
}
 \cite{jesus2019thesis,morelli2021,hanamura2021,boeyens2021noninformative}
\begin{equation}
    \bar{\epsilon}_{\boldsymbol{y}
} (x) = \int d\theta\,p(\theta|x, \boldsymbol{y}
)\,\mathcal{D}[\tilde{\theta}_{\boldsymbol{y}
}(x),\theta],
\label{eq:err-exp}
\end{equation}
for some deviation function $ \mathcal{D}$. 
In this way, the unknown parameter $\Theta$ can, for fixed estimator and POM, be assigned a single empirical number which quantifies its value.\footnote{
Nevertheless, note that failing to consider the full posterior density [Eq.~\eqref{eq:posterior}] might leave potentially valuable information unused \cite{toussaint2011}.
}
We thus need a procedure to identify which estimator best captures the essential information of Eq.~\eqref{eq:posterior}.

Since such a task is to be carried out without knowledge of a specific measurement outcome, one proceeds by demanding the overall uncertainty 
\begin{equation}
    \bar{\epsilon}_{\boldsymbol{y}
} = \int dx\,p(x|\boldsymbol{y}
)\,\bar{\epsilon}_{\boldsymbol{y}}(x),
    \label{eq:overall-err}
\end{equation}
as opposed to using just $\bar{\epsilon}_{\boldsymbol{y}
}(x)$, to be minimal \cite{helstrom1976,jaynes2003, rafal2015, demkowicz2020}, where $p(x|\boldsymbol{y}
) = \int d\theta\,p(\theta) p(x|\theta, \boldsymbol{y}
)$. 
This implies that the notion of `optimal strategy' crucially depends on the chosen measure of uncertainty \cite{bialynicki1993, alfredo2013}, thus reinforcing the idea that deviation functions must be carefully justified.

The fact that $\Theta$ is a scale parameter leads to the logarithmic deviation function $\mathcal{D}(\tilde{\theta},\theta) = \log^2(\tilde{\theta}/ \theta)$ \cite{rubio2020}.
This is a type of relative error which is suitable for scale estimation as it satisfies the following properties: symmetry, $\mathcal{D}(\tilde{\theta},\theta) = \mathcal{D}(\theta, \tilde{\theta})$; scale invariance, $\mathcal{D}(\gamma \tilde{\theta},\gamma \theta) = \mathcal{D}(\tilde{\theta},\theta)$; and having its absolute minimum at $\tilde{\theta} = \theta$, from (towards) where it grows (decreases) monotonically when $\tilde{\theta} > \theta$ ($\tilde{\theta} < \theta$). 
Its derivation, available in Ref.~\cite{rubio2020}, will be revisited in Sec.~\ref{sec:multiparameter} while addressing the multiparameter case.
By inserting $\mathcal{D}(\tilde{\theta},\theta) = \log^2(\tilde{\theta}/ \theta)$ into Eq.~\eqref{eq:overall-err}, and using the identity $p(x|\boldsymbol{y})\,p(\theta|x, \boldsymbol{y}) = p(x,\theta|\boldsymbol{y})$, one finds the mean logarithmic error 
\begin{equation} 
\bar{\epsilon}_{\boldsymbol{y}} \mapsto \bar{\epsilon}_{\boldsymbol{y}, \mathrm{mle}} = \int dx\,d\theta \hspace{0.2em} p(x, \theta | \boldsymbol{y}) \log^2\left[\frac{\tilde{\theta}_{\boldsymbol{y}}(x)}{\theta}\right];
\label{eq:th-mle}
\end{equation}
this is the uncertainty quantifier that we will use. 

If an experimental platform is already built, the associated joint probability $p(x, \theta | \boldsymbol{y})$ can be fixed, so that the search for a good estimate amounts to minimising Eq.~\eqref{eq:th-mle} with respect to $\tilde{\theta}_{\boldsymbol{y}}(x)$ \cite{rubio2020}. 
However, quantum-mechanical probabilities
 allow for this minimisation to be performed while the experimental design is itself optimised. 
Specifically, using $p(x,\theta|\boldsymbol{y}) = p(\theta)\,p(x|\theta, \boldsymbol{y})$, together with the first equality in Eq.~\eqref{eq:likelihood-construction}, we can rewrite Eq.~\eqref{eq:th-mle} as 
\begin{equation}
    \bar{\epsilon}_{\boldsymbol{y}, \mathrm{mle}} = \mathrm{Tr}\left\lbrace\int dx\,M_{\boldsymbol{y}}(x)\, W_{\boldsymbol{y}}[\tilde{\theta}_{\boldsymbol{y}}(x)]\right\rbrace,
\label{eq:mle_quantum}
\end{equation}
where the operator $W_{\boldsymbol{y}}[\tilde{\theta}_{\boldsymbol{y}}(x)]$ reads \cite{helstrom1976}
\begin{equation}
    W_{\boldsymbol{y}}[\tilde{\theta}_{\boldsymbol{y}}(x)] = \int d\theta\,p(\theta)\,\rho_{\boldsymbol{y}}(\theta)\log^2\left[\frac{\tilde{\theta}_{\boldsymbol{y}}(x)}{\theta}\right].
    \label{eq:aux_W}
\end{equation}
Provided that $p(\theta)$ and $\rho_{\boldsymbol{y}}(\theta)$ are known, we can then search for the estimator \emph{and} the POM that together minimise Eq.~\eqref{eq:mle_quantum}. 
In other words, we seek the solution to the optimisation problem
\begin{equation}
   \min_{M_{\boldsymbol{y}}(x),\,\tilde{\theta}_{\boldsymbol{y}}(x)} 
   \mathrm{Tr}\left\lbrace\int dx\,M_{\boldsymbol{y}}(x)\, W_{\boldsymbol{y}}[\tilde{\theta}_{\boldsymbol{y}}(x)]\right\rbrace
   = \bar{\epsilon}_{\boldsymbol{y}, \mathrm{min}}.
   \label{eq:min_def}
\end{equation}
Once found, the optimal POM informs how the measurement protocol should be designed as to extract maximum information about $\Theta$, while the associated estimator provides a rule to optimally process measurement data into an estimate. 

\section{Quantum estimation of scale parameters} \label{sec:main_result}

\subsection{Fundamental minimum}\label{sec:min-derivation}

Our first task is to calculate the minimum in Eq.~\eqref{eq:min_def}. 
Start by expanding $W_{\boldsymbol{y}}[\tilde{\theta}_{\boldsymbol{y}}(x)]$ in Eq.~\eqref{eq:aux_W} as
\begin{align}
    W_{\boldsymbol{y}}[\tilde{\theta}_{\boldsymbol{y}}(x)] = &  \varrho_{\boldsymbol{y},2} 
    + \varrho_{\boldsymbol{y},0} \log^2\left[\frac{\tilde{\theta}_{\boldsymbol{y}}(x)}{\theta_u}\right] 
	\nonumber \\    
    &- 2\varrho_{\boldsymbol{y},1} \log\left[\frac{\tilde{\theta}_{\boldsymbol{y}}(x)}{\theta_u}\right],
    \label{eq:W_proof}
\end{align}
where the operator $\varrho_{\boldsymbol{y},k}$ is defined as 
\begin{equation}
    \varrho_{\boldsymbol{y},k} \coloneqq \int d\theta \,p(\theta) \rho_{\boldsymbol{y}}(\theta) \log^k\left(\frac{\theta}{\theta_u}\right)
    \label{eq:quantum_moments}
\end{equation}
and $\theta_u$ is an arbitrary constant, with the same units as $\theta$, introduced to guarantee that the argument of logarithmic functions is dimensionless \cite{matta2011}, i.e.,
\begin{equation}
    \log\left[\frac{\tilde{\theta}_{\boldsymbol{y}}(x)}{\theta}\right] = \log\left[\frac{\tilde{\theta}_{\boldsymbol{y}}(x)}{\theta_u}\right] - \log\left(\frac{\theta}{\theta_u}\right).
\end{equation}
Inserting Eq.~\eqref{eq:W_proof} into Eq.~\eqref{eq:mle_quantum} further gives 
\begin{align}
    \bar{\epsilon}_{\boldsymbol{y}, \mathrm{mle}} = & \int d\theta\,p(\theta)\log^2\left(\frac{\theta}{\theta_u}\right)
	\nonumber \\    
    & + \mathrm{Tr}(\varrho_{\boldsymbol{y},0} \mathcal{A}_{\boldsymbol{y},2} - 2\varrho_{\boldsymbol{y},1} \mathcal{A}_{\boldsymbol{y},1}),
    \label{eq:mle_optimisation_in_progress}
\end{align}
where we have introduced the operator 
\begin{equation}
    \mathcal{A}_{\boldsymbol{y},k} \coloneqq \int dx\,M_{\boldsymbol{y}}(x) \log^k\left[\frac{\tilde{\theta}_{\boldsymbol{y}}(x)}{\theta_u}\right].
    \label{eq:Aop-aux}
\end{equation}
Since the first term of $\bar{\epsilon}_{\boldsymbol{y}, \mathrm{mle}}$ in Eq.~\eqref{eq:mle_optimisation_in_progress} is estimator- and POM-independent, we only need to minimise the trace term.

Next, define a new estimator function $\tilde{\omega}_{\boldsymbol{y}}(x)$ as
\begin{equation}
    \tilde{\omega}_{\boldsymbol{y}}(x) \coloneqq \log\left[\frac{\tilde{\theta}_{\boldsymbol{y}}(x)}{\theta_u}\right],
    \label{eq:omega-aux}
\end{equation}
so that Eq.~\eqref{eq:Aop-aux} becomes
\begin{equation}
    \mathcal{A}_{\boldsymbol{y},k} = \int dx\,M_{\boldsymbol{y}}(x)\,\tilde{\omega}_{\boldsymbol{y}}(x)^k.
    \label{eq:Aop-omega}
\end{equation}
One then sees that
\begin{align}
    \mathcal{A}_{\boldsymbol{y},2} - \mathcal{A}_{\boldsymbol{y},1}^2 
    = & \int dx\,M_{\boldsymbol{y}}(x)\,\tilde{\omega}_{\boldsymbol{y}}(x)^2 
	\nonumber \\   
    & - \left[\int dx\,M_{\boldsymbol{y}}(x)\,\tilde{\omega}_{\boldsymbol{y}}(x)\right]^2 
    \geq 0
    \label{eq:jensen_op}
\end{align}
due to Jensen's operator inequality \cite{hansen2003,macieszczak2014bayesian}, which in turn implies the inequality
\begin{equation}
    \mathrm{Tr}(\varrho_{\boldsymbol{y},0} \mathcal{A}_{\boldsymbol{y},2} - 2\varrho_{\boldsymbol{y},1} \mathcal{A}_{\boldsymbol{y},1}) \geq \mathrm{Tr}(\varrho_{\boldsymbol{y},0} \mathcal{A}_{\boldsymbol{y},1}^2 - 2\varrho_{\boldsymbol{y},1} \mathcal{A}_{\boldsymbol{y},1})
    \label{eq:aux_inequality}
\end{equation}
for the trace in Eq.~\eqref{eq:mle_optimisation_in_progress}.
The inequality in Eq.~\eqref{eq:jensen_op}---and thus that in Eq.~\eqref{eq:aux_inequality}---is saturated when the POM is projective, i.e., 
\begin{equation}
    M_{\boldsymbol{y}}(x) = \mathcal{P}_{\boldsymbol{y}}(x),
    \label{eq:pom-condition}
\end{equation}
with $\mathcal{P}_{\boldsymbol{y}}(x)\,\mathcal{P}_{\boldsymbol{y}}(x') \mapsto \delta(x-x')\,\mathcal{P}_{\boldsymbol{y}}(x')$.
Therefore, we can restrict our calculation to the set of projective measurements without loss of optimality.\footnote{
An analogous argument for the minimisation of the mean square error has been considered before \cite[Appendix~A]{macieszczak2014bayesian}.}

From these considerations we see that minimising the mean logarithmic error $\bar{\epsilon}_{\boldsymbol{y},\mathrm{mle}}$ amounts to minimising the right hand side of Eq.~\eqref{eq:aux_inequality}.
Given that both the estimator and the POM appear inside the operator $\mathcal{A}_{\boldsymbol{y}, 1}$, we can proceed as
\begin{equation}
    \min_{\mathcal{A}_{\boldsymbol{y}, 1}} \mathrm{Tr}(\varrho_{\boldsymbol{y},0} \mathcal{A}_{1, \boldsymbol{y}}^2 - 2\varrho_{\boldsymbol{y}, 1} \mathcal{A}_{\boldsymbol{y}, 1}) = - \mathrm{Tr}(\varrho_{\boldsymbol{y}, 0} \mathcal{S}_{\boldsymbol{y}}^2),
    \label{eq:partial_min}
\end{equation}
where the operator $\mathcal{S}_{\boldsymbol{y}}$ is solution to the Lyaponuv equation
\begin{equation}
    \mathcal{S}_{\boldsymbol{y}}\varrho_{\boldsymbol{y},0}+\varrho_{\boldsymbol{y},0}\mathcal{S}_{\boldsymbol{y}} = 2\varrho_{\boldsymbol{y},1}
    \label{eq:lyaponuv_eq}
\end{equation}
and the minimum is achieved when $\mathcal{A}_{\boldsymbol{y},1} = \mathcal{S}_{\boldsymbol{y}}$.
The details of this calculation, which is based on the calculus of variations in operator form \cite{personick1971,mathematics2004,bernad2018}, can be found in 
Appendix~\ref{app:minimisation_aux}.

By combining Eq.~\eqref{eq:mle_optimisation_in_progress} with the equality in Eq.~\eqref{eq:aux_inequality} and the minimum in Eq.~\eqref{eq:partial_min}, we finally arrive at the minimum mean logarithmic error
\begin{equation}
    \bar{\epsilon}_{\boldsymbol{y}, \mathrm{min}} = \int d\theta\,p(\theta)\log^2\left(\frac{\theta}{\theta_u}\right) - \mathrm{Tr}(\varrho_{\boldsymbol{y},0} \mathcal{S}_{\boldsymbol{y}}^2).
    \label{eq:quantum_minimum}
\end{equation}
The practical importance of this result is apparent: for given prior density $p(\theta)$ and density operator $\rho_{\boldsymbol{y}}(\theta)$, it provides the means to calculate fundamental limits to the precision in quantum scale estimation.\footnote{
Note that since $\theta_u$ does not appear in Eq.~\eqref{eq:mle_quantum}, subsequent manipulations of the mean logarithmic error---including the minimum in Eq.~\eqref{eq:quantum_minimum}---are also independent of its specific value.
}

\subsection{Optimal quantum strategy}\label{sec:opt-straegy}

The next step is to identify the estimator function and the POM reaching the minimum in Eq.~\eqref{eq:quantum_minimum}. 
That is, we seek the optimal quantum strategy. 
To find it, first recall that Eq.~\eqref{eq:quantum_minimum} relies on fulfilling Eq.~\eqref{eq:pom-condition} and the condition $\mathcal{A}_{\boldsymbol{y},1} = \mathcal{S}_{\boldsymbol{y}}$.
Using Eq.~\eqref{eq:Aop-omega} and the eigendecomposition $\mathcal{S}_{\boldsymbol{y}} = \int ds\,\mathcal{P}_{\boldsymbol{y}}(s)\,s$, these can be combined into a single condition as
\begin{equation}
    \int dx\,\mathcal{P}_{\boldsymbol{y}}(x)\,\tilde{\omega}_{\boldsymbol{y}}(x) = \int ds\,\mathcal{P}_{\boldsymbol{y}}(s)\,s,
    \label{eq:unified-condition}
\end{equation}
which leads to $\mathcal{P}_{\boldsymbol{y}}(x)\,dx = \mathcal{P}_{\boldsymbol{y}}(s)\,ds$ and $\tilde{\omega}_{\boldsymbol{y}}(x = s) = s$.

According to the first of these constraints, the optimal POM is given by the projectors associated with $\mathcal{S}_{\boldsymbol{y}}$, i.e., 
\begin{equation}
    M_{\boldsymbol{y}}(x) = \mathcal{P}_{\boldsymbol{y}}(x) \mapsto \mathcal{M}_{\boldsymbol{y}}(s) = \mathcal{P}_{\boldsymbol{y}}(s).
    \label{eq:opt-pom-final}
\end{equation}
To identify the meaning of the second constraint, we need to revert the transformation in Eq.~\eqref{eq:omega-aux}, i.e.,
\begin{equation}
    \tilde{\theta}_{\boldsymbol{y}}(x) = \theta_u\,\mathrm{exp}\left[ \tilde{\omega}_{\boldsymbol{y}}(x) \right].
\end{equation}
The optimal estimator function is then found to be
\begin{equation}
    \tilde{\theta}_{\boldsymbol{y}}(x) \mapsto \tilde{\vartheta}_{\boldsymbol{y}}(s) = \theta_u \exp(s).
    \label{eq:opt-est-final}
\end{equation}

Two consistency checks are in order. First, one may confirm [Appendix~\ref{app:HH-path}] that Eqs.~\eqref{eq:opt-pom-final} and \eqref{eq:opt-est-final} satisfy the conditions for the optimal quantum strategy laid out by Holevo \cite{holevo1973,holevo1973b} and Helstrom \cite{helstrom1974,helstrom1976}.

Secondly, Eq.~\eqref{eq:opt-est-final} can be shown to recover the probability-based version of the optimal estimator that was derived in Ref.~\cite{rubio2020}, as follows.
Consider the quantity $\mathrm{Tr}[\mathcal{P}_{\boldsymbol{y}}(s)\varrho_{\boldsymbol{y},1}]$ and insert $\varrho_{\boldsymbol{y},1} = (\mathcal{S}_{\boldsymbol{y}}\varrho_{\boldsymbol{y},0} + \varrho_{\boldsymbol{y},0}\mathcal{S}_{\boldsymbol{y}})/2$, which is true by virtue of Eq.~\eqref{eq:lyaponuv_eq}; then 
\begin{align}
    \mathrm{Tr}[\mathcal{P}_{\boldsymbol{y}}(s)\varrho_{\boldsymbol{y},1}] & = \mathrm{Re}\lbrace\mathrm{Tr}[\mathcal{P}_{\boldsymbol{y}}(s)\,\mathcal{S}_{\boldsymbol{y}}\varrho_{\boldsymbol{y},0}]\rbrace
    \nonumber \\
    & = \int d\breve{s}\,\breve{s}\, \mathrm{Re}\lbrace\mathrm{Tr}[\mathcal{P}_{\boldsymbol{y}}(s)\mathcal{P}_{\boldsymbol{y}}(\breve{s}) \varrho_{\boldsymbol{y},0}]\rbrace
    \nonumber \\
    & = \int d\breve{s}\,\breve{s}\,\delta(s - \breve{s})\,\mathrm{Tr}[\mathcal{P}_{\boldsymbol{y}}(\breve{s}) \varrho_{\boldsymbol{y},0} ]
    \nonumber \\
    & = s\,\mathrm{Tr}[\mathcal{P}_{\boldsymbol{y}}(s) \varrho_{\boldsymbol{y},0}],
\end{align}
so that
\begin{equation}
    s = \frac{\mathrm{Tr}[\mathcal{P}_{\boldsymbol{y}}(s)\varrho_{\boldsymbol{y},1}]}{\mathrm{Tr}[\mathcal{P}_{\boldsymbol{y}}(s)\varrho_{\boldsymbol{y},0}]}. 
    \label{eq:gqt-qse_equivalence}
\end{equation}
By introducing the expressions for $\rho_{\boldsymbol{y}, 0}$ and $\rho_{\boldsymbol{y},1}$ [Eq.~\eqref{eq:quantum_moments}] in Eq.~\eqref{eq:gqt-qse_equivalence}, 
we see that 
\begin{equation}
\mathrm{Tr}[\mathcal{P}_{\boldsymbol{y}}(s)\varrho_{\boldsymbol{y},0}] 
= \int d\theta\, p(\theta)\, \mathrm{Tr}[\mathcal{P}_{\boldsymbol{y}}(s) \rho_{\boldsymbol{y}}(\theta)] = p(s | \boldsymbol{y})
\label{eq:opt-evidence}
\end{equation}
and 
\begin{align}
    s &= \frac{1}{p(s | \boldsymbol{y})}\int d\theta\, p(\theta)\, \mathrm{Tr}[\mathcal{P}_{\boldsymbol{y}}(s) \rho_{\boldsymbol{y}}(\theta)] \log\left(\frac{\theta}{\theta_u}\right)
\nonumber \\ 
	&= \int d\theta\, \frac{p(\theta)\, p(s | \theta, \boldsymbol{y})}{p(s | \boldsymbol{y})} \log\left(\frac{\theta}{\theta_u}\right)
	\nonumber \\    
     &= \int d\theta\, p(\theta|s, \boldsymbol{y})\log\left(\frac{\theta}{\theta_u}\right).
\end{align}
Finally, using Eq.~\eqref{eq:opt-est-final} leads to
\begin{align}
    \tilde{\vartheta}_{\boldsymbol{y}}(s) =
    \theta_u \exp\left[\int d\theta\, p(\theta|s, \boldsymbol{y})\log\left(\frac{\theta}{\theta_u}\right)\right],
    \label{eq:optest-probability}
\end{align}
in agreement with the result in Ref.~\cite{rubio2020}.  

We have thus identified a transparent and straightforward procedure---solving Eq.~\eqref{eq:lyaponuv_eq} for $\mathcal{S}_{\boldsymbol{y}}$---to construct the optimal POM [Eq.\eqref{eq:opt-pom-final}] and estimator [Eqs.~\eqref{eq:opt-est-final} and \eqref{eq:optest-probability}] for any prior probability and density operator, when the error is logarithmic and squared.
This, together with the minimum in Eq.~\eqref{eq:quantum_minimum}, fully solves the optimisation problem in Eq.~\eqref{eq:min_def}. 
For an alternative derivation using the method in Ref.~\cite{jesus2020mar}, see
Appendix~\ref{app:JJ-path}.

\section{Examples}\label{sec:example}

Two applications of quantum scale estimation are illustrated in the following: thermometry, and the estimation of the lifetime of an atomic state.

\subsection{Global quantum thermometry}\label{sec:thermo-example}

In Ref.~\cite{rubio2020}, global quantum thermometry was formulated for a given likelihood model. 
Within this framework, optimising the design of a thermometric experiment can only be achieved by assessing, one by one, the uncertainty associated with different likelihood functions. 
However, such a procedure does not guarantee that the optimal design will be found.
To address this problem, one would ideally optimise the estimation error with respect to either the state, the POM, or both.
In Secs.~\ref{sec:thermal-states} - \ref{sec:energy-spectrum}, quantum scale estimation is shown to enable such a possibility for Bayesian thermometry.
Additionally, even if the likelihood model is assumed at the outset, the formalism in this work still provides a useful language to describe different thermometric protocols in a unified way; we will see this in Sec.~\ref{sec:other-laws}.
The following example thus demonstrates that quantum scale estimation, when applied to thermometry, generalises the framework in Ref.~\cite{rubio2020}.

\subsubsection{Probability law for thermal states}\label{sec:thermal-states}

Let $H = \sum_n \ketbra{n} \varepsilon_n$ be the Hamiltonian for a quantum probe---the `thermometer'---which is in weak thermal contact with another system whose temperature $\Theta = T$ we wish to estimate \cite{correa2015, mehboudi2019}. 
When the probe is fully equilibrated, it can be described by a thermal state
\begin{equation}
    \rho_{\boldsymbol{y}}(\theta)  = \frac{\mathrm{exp}[-H/(k_B \theta)]}{\mathrm{Tr}\lbrace \mathrm{exp}[-H/(k_B \theta)] \rbrace}
	= \sum_n \ketbra{n} h_n\hspace{-0.1em}\left(\frac{\boldsymbol{y}}{\theta}\right), 
    \label{eq:thermal_state}
\end{equation}
where $k_B$ is Boltzmann's constant, 
\begin{equation}
h_n\hspace{-0.1em}\left(\frac{\boldsymbol{y}}{\theta}\right) = \frac{\mathrm{exp}(- y_n/\theta)}{\sum_{m}  \mathrm{exp}(-y_m/\theta)},
\label{eq:state-components}
\end{equation}
and $\boldsymbol{y} = (\varepsilon_0, \varepsilon_1, \dots)/k_B$ is a vector of temperatures.
That is, the energy spectrum gives rise to the set of known parameters. 

Consider, in addition, an arbitrary POM
\begin{equation}
M_{\boldsymbol{y}}(x) = \sum_{n m} \ketbra{n}{m} M_{\boldsymbol{y}, n m}(x)
\end{equation}
where both $M_{\boldsymbol{y}, n m}(x) = \bra{n} M_{\boldsymbol{y}}(x)\ket{m}$ and $x$ are dimensionless.
Then, the Born rule reads
\begin{equation}
p(x| \boldsymbol{y},\theta) = \sum_{n} M_{\boldsymbol{y}, n n}(x)\,h_n\hspace{-0.1em}\left(\frac{\boldsymbol{y}}{\theta}\right).
\label{eq:bornrule-example}
\end{equation}

Here, $h_n$ is trivially invariant under transformations $\lbrace \boldsymbol{y}' \mapsto \gamma \boldsymbol{y},\,\theta' \mapsto \gamma \theta \rbrace$. 
But the same must hold true for $M_{\boldsymbol{y}, n n}$ due to dimensional consistency, since $\boldsymbol{y}$ is dimensioned. 
Consequently, $p(x| \boldsymbol{y},\theta) = h(x|\boldsymbol{y}/\theta)$ [Sec.~\ref{sec:formulation}], and so the formalism for quantum scale estimation in Sec.~\ref{sec:main_result} applies.

\subsubsection{Measurement strategy}\label{sec:thermal-pom}

For an arbitrary prior density $p(\theta)$, we wish to find the POM that is optimal given the state in Eq.~\eqref{eq:thermal_state}. 
The first step is to insert Eqs.~\eqref{eq:thermal_state} and \eqref{eq:state-components} into Eq.~\eqref{eq:quantum_moments}, which gives
\begin{equation}
    \varrho_{\boldsymbol{y}, k} = \sum_{n}\ketbra{n}\chi_{\boldsymbol{y}}^{n,k},
    \label{eq:quantum_moments_GQT}
\end{equation}
with each coefficient $\chi_{\boldsymbol{y}}^{n,k}$ defined as
\begin{equation}
    \chi_{\boldsymbol{y}}^{n,k} = \int d\theta\,p(\theta)\,\frac{\exp(-y_n/\theta) \log^k(\theta/\theta_u)}{\sum_m \exp(-y_m/\theta)}.
    \label{eq:B_def}
\end{equation}

Next, using $\mathcal{S}_{\boldsymbol{y}} = \sum_{nm} \ketbra{n}{m} \mathcal{S}_{\boldsymbol{y}, nm}$, where $\mathcal{S}_{\boldsymbol{y}, nm} = \bra{n} \mathcal{S}_{\boldsymbol{y}} \ket{m}$, and taking $\varrho_{\boldsymbol{y},0}$ and $\varrho_{\boldsymbol{y},1}$ from Eq.~\eqref{eq:quantum_moments_GQT}, the Lyaponuv equation \eqref{eq:lyaponuv_eq} becomes
\begin{equation}
    \sum_{nm} \ketbra{n}{m} [(\chi_{\boldsymbol{y}}^{n,0}+\chi_{\boldsymbol{y}}^{m,0})\, \mathcal{S}_{\boldsymbol{y}, nm} - 2 \chi_{\boldsymbol{y}}^{n,1}\delta_{nm}] = 0.
\end{equation}
Solving for each component $\mathcal{S}_{\boldsymbol{y}, n m}$, we find
\begin{equation}
    \mathcal{S}_{\boldsymbol{y}} = \sum_n \ketbra{n}\frac{\chi_{\boldsymbol{y}}^{n,1}}{\chi_{\boldsymbol{y}}^{n,0}},
    \label{eq:S_operator_GQT}
\end{equation}
which is diagonal in the energy basis $\lbrace \ketbra{n}{n} \rbrace$. 
We can further rewrite Eq.~\eqref{eq:S_operator_GQT} in the language of continuous variables as $\mathcal{S}_{\boldsymbol{y}} = \int ds\,\mathcal{P}_{\boldsymbol{y}}(s)\,s$ by using the projective measurement
\begin{equation}
    \mathcal{P}_{\boldsymbol{y}}(s) = \sum_n \ketbra{n} \delta (s - s_{\boldsymbol{y},n}),
\end{equation}
where we have defined $s_{\boldsymbol{y},n}  \coloneqq \chi_{\boldsymbol{y}}^{n,1}/\chi_{\boldsymbol{y}}^{n,0}$.
Hence, the optimal POM is $M_{\boldsymbol{y}}(x) \mapsto \mathcal{M}_{\boldsymbol{y}}(s) =  \mathcal{P}_{\boldsymbol{y}}(s)$.

Let us now examine the probability density associated with this POM. 
By inserting the latter into Eq.~\eqref{eq:bornrule-example}, 
and noticing that $h_n(\boldsymbol{y}/\theta) = p(n|\theta,\boldsymbol{y})$ is a discrete probability, we have 
\begin{equation}
p(s|\theta,\boldsymbol{y}) = \sum_n \delta(s - s_{\boldsymbol{y},n}) \,p(n|\theta,\boldsymbol{y}).
\label{eq:gqt-optstatistics}
\end{equation}
Then, the optimal strategy effectively consists in measuring energy.
Note that Eq.~\eqref{eq:gqt-optstatistics} is invariant under changes in the scale of temperature---since both $s_{\boldsymbol{y},n}$ and $p(n|\theta,\boldsymbol{y})$ are---in consistency with the initial premise. 

From a practical perspective, this result teaches us two important lessons. 
First, it demonstrates that $\mathcal{S}_{\boldsymbol{y}}$ can give rise to a realisable measurement protocol---in this case, energy measurements. 
This is crucial as it implies that quantum scale estimation will be applicable at least in some experiments. 

Secondly, this is valid `globally' \cite{paris2009}, in the sense that it holds for an arbitrarily large prior temperature range. 
In contrast, `local' thermometry assumes---often implicitly---a prior temperature range $[\theta_{\mathrm{min}}, \theta_{\mathrm{max}}]$ such that $\theta_{\mathrm{max}}/\theta_{\mathrm{min}} \sim 1$. 
Performing energy measurements on the state in Eq.~\eqref{eq:thermal_state} was known to be optimal in local estimation \cite{stace2010quantum, DePasquale2018, mehboudi2019}; our discussion shows this to hold \emph{also} in scale-invariant global thermometry \cite{rubio2020}, which was an open question. 

\subsubsection{Estimate and experimental error}\label{sec:example-est-err}

We next identify the optimal estimator.
From Eq.~\eqref{eq:optest-probability},
\begin{equation}
p(s|\boldsymbol{y})\, \mathrm{log}\left[ \frac{\tilde{\vartheta}_{\boldsymbol{y}}(s)}{\theta_u}\right] = \int d\theta\, p(\theta)\,p(s|\theta,\boldsymbol{y}) \, \mathrm{log}\left( \frac{\theta}{\theta_u} \right).
\label{eq:discrete-est-aux1}
\end{equation}
Using Eq.~\eqref{eq:gqt-optstatistics}, we see that 
\begin{align}
	p(s|\boldsymbol{y}) &= \int d\theta\,p(\theta)\,p(s|\theta, \boldsymbol{y}) 
		\nonumber \\
	&= \sum_n  \delta(s-s_{\boldsymbol{y},n}) \int d\theta\,p(\theta)\,p(n|\theta, \boldsymbol{y}) 
		\nonumber \\
	&= \sum_n \delta(s-s_{\boldsymbol{y},n})\,p(n|\boldsymbol{y}).
	\label{eq:discrete-est-aux2}
\end{align}
Inserting Eqs.~\eqref{eq:gqt-optstatistics} and \eqref{eq:discrete-est-aux2} into Eq.~\eqref{eq:discrete-est-aux1}, integrating over $s$, and rearranging, we get $\sum_n  F_n = 0$, where
\begin{align}
F_n \coloneqq & \, p(n|\boldsymbol{y})\, \mathrm{log}\left[ \frac{\tilde{\vartheta}_{\boldsymbol{y}}(s_{\boldsymbol{y},n})}{\theta_u}\right]
\nonumber \\
	& - \int d\theta\, p(\theta)\,p(n|\theta, \boldsymbol{y})\,\mathrm{log}\left( \frac{\theta}{\theta_u} \right).
\end{align}
Taking each addend as $F_n = 0$, and solving for $\tilde{\vartheta}_{\boldsymbol{y}}(s_{\boldsymbol{y},n})$, finally renders
\begin{equation}
\tilde{\vartheta}_{\boldsymbol{y}}(s_{\boldsymbol{y},n}) = \theta_u \,\mathrm{exp}\left[\int d\theta\,p(\theta|n,\boldsymbol{y})\,\mathrm{log}\left( \frac{\theta}{\theta_u} \right)\right],
\label{eq:opt-est-example}
\end{equation}
where $p(\theta|n,\boldsymbol{y}) \propto p(\theta)\,\mathrm{exp}(-y_n/\theta)/[\sum_m\mathrm{exp}(-y_m/\theta)]$.
This is the optimal estimator for thermal quantum states. 

To calculate the measurement-dependent error associated with this estimator, we first introduce the optimal POM $\mathcal{P}_{\boldsymbol{y}}(s)$ and the deviation function $ \mathcal{D}(\tilde{\theta},\theta) = \mathrm{log}^2(\tilde{\theta}/\theta)$ in Eq.~\eqref{eq:err-exp}, that is,
\begin{equation}
\bar{\epsilon}_{\boldsymbol{y}, \mathrm{mle}}(s) = \int d\theta\,p(\theta|s, \boldsymbol{y})\,\mathrm{log}^2 \left[ \frac{\tilde{\vartheta}_{\boldsymbol{y}}(s)}{\theta} \right].
\end{equation}
Next, we rewrite this expression as
\begin{equation}
p(s|\boldsymbol{y})\,\bar{\epsilon}_{\boldsymbol{y}, \mathrm{mle}}(s) = \int d\theta\,p(\theta)\,p(s|\theta, \boldsymbol{y})\,\mathrm{log}^2 \left[ \frac{\tilde{\vartheta}_{\boldsymbol{y}}(s)}{\theta} \right].
\end{equation}
We can then proceed, \emph{mutatis mutandis}, as with Eq.~\eqref{eq:discrete-est-aux1}, finding 
\begin{equation}
	\bar{\epsilon}_{\boldsymbol{y}, \mathrm{mle}}(s_{\boldsymbol{y}, n}) =  \int d\theta\,p(\theta|n, \boldsymbol{y})\,\mathrm{log}^2 \left[ \frac{\tilde{\vartheta}_{\boldsymbol{y}}(s_{\boldsymbol{y}, n})}{\theta} \right].
\end{equation}
By inserting Eq.~\eqref{eq:opt-est-example}, we finally arrive at 
\begin{align}
	\bar{\epsilon}_{\boldsymbol{y}, \mathrm{mle}}(s_{\boldsymbol{y}, n}) 
	= & \int d\theta\,p(\theta|n,\boldsymbol{y})\,\mathrm{log}^2 \left( \frac{\theta}{\theta_u} \right) 
	\nonumber \\	
	&- \left[ \int d\theta\,p(\theta|n,\boldsymbol{y})\,\mathrm{log} \left( \frac{\theta}{\theta_u} \right) \right]^2.
	\label{eq:opt-exp-err-example}
\end{align}

Given a thermometric experiment described by Eq.~\eqref{eq:thermal_state}---even if only effectively---Eqs.~\eqref{eq:opt-est-example} and \eqref{eq:opt-exp-err-example} allow us to report optimal temperature estimates as
\begin{equation}
\tilde{\vartheta}_{\boldsymbol{y}}(s_{\boldsymbol{y},n}) \pm \Delta \tilde{\vartheta}_{\boldsymbol{y}}(s_{\boldsymbol{y},n}),
\end{equation}
where $\Delta \tilde{\vartheta}_{\boldsymbol{y}}(s_{\boldsymbol{y},n}) \coloneqq \tilde{\vartheta}_{\boldsymbol{y}}(s_{\boldsymbol{y},n}) \sqrt{\bar{\epsilon}_{\boldsymbol{y}, \mathrm{mle}}(s_{\boldsymbol{y}, n}) }$ plays the role of an error bar \cite{rubio2020}.

\subsubsection{Theoretical optimum and energy spectrum}\label{sec:energy-spectrum}

We have so far assumed that the energy spectrum in Eq.~\eqref{eq:thermal_state} is given \emph{a priori}. 
However, one may also be interested in searching for the energy spectrum which leads to the greatest precision on average.
In local thermometry \cite{correa2015, mehboudi2019}, this is done by maximising the quantum Fisher information \cite{paris2009, rafal2015}. 
Here, the task is instead to minimise Eq.~\eqref{eq:quantum_minimum} over $\boldsymbol{y}$.
The key advantage of this approach is the possibility of including prior information---via $p(\theta)$---in an exact and controlled fashion. 

By inserting Eq.~\eqref{eq:S_operator_GQT} into Eq.~\eqref{eq:quantum_minimum}, we have
\begin{equation}
    \bar{\epsilon}_{\boldsymbol{y}, \mathrm{min}} 
    = \int d\theta\,p(\theta)\log^2\left(\frac{\theta}{\theta_u}\right) 
    - \sum_n \frac{\left(\chi_{\boldsymbol{y}}^{n,1}\right)^2}{\chi_{\boldsymbol{y}}^{n,0}}.
\end{equation}
Since the first term is spectrum-independent, minimising this uncertainty amounts to maximising the quantity
\begin{equation}
\sum_n \frac{\left(\chi_{\boldsymbol{y}}^{n,1}\right)^2}{\chi_{\boldsymbol{y}}^{n,0}}
\label{eq:maximisation-condition}
\end{equation}
over $\boldsymbol{y}$, for given $p(\theta)$. 

According to Eq.~\eqref{eq:maximisation-condition}, the form of the optimal energy spectrum will generally depend on the prior temperature range. 
One may thus expect an emergence of different phases for the associated probes as the temperature range is increased---this was first noted by \citet{mok2020} by using a global version of the Cram\'{e}r-Rao bound \cite{pearce2017}. 
Given that quantum scale estimation offers a more accurate control of which prior information enters our temperature estimates, the search of optimal energy spectra for thermal states should be revisited via Eq.~\eqref{eq:maximisation-condition}. 
This is left for future work.

\subsubsection{Other probability laws in thermometry}
\label{sec:other-laws}

Previous sections have relied on the class of probability models in Eq.~\eqref{eq:posterior}.
We have seen that Eq.~\eqref{eq:posterior} emerges from assumptions (i - iii) in Sec.~\ref{sec:formulation}, and that these provide one way of formulating scale estimation at the level of quantum operators.
Nevertheless, in cases where working with probability functions suffices---e.g., when both the state and the measurement have already been chosen in a given experiment---one can relax these assumptions to find other scale-invariant laws in thermometry.  
For the sake completeness, this is now illustrated.
The functional equations in this section are solved as indicated in Ref.~\cite[Ch.~3]{aczel1987}.

Suppose we admit assumption (i), so that the prior probability is still constructed as in Sec.~\ref{sec:formulation}. 
We also keep assumption (ii) for a single parameter $y$, but we drop assumption (iii). 
Unlike in Sec.~\ref{sec:formulation}, here we only need to impose $p(x|\theta, y) = p(x|\gamma \theta, \gamma y)$, which leads to 
\begin{equation}
p(x|\theta, y) = h\hspace{-0.1em}\left(x, \frac{y}{\theta} \right),
\label{eq:poisson-family}
\end{equation}
where $h$ is a free function. 
Eq.~\eqref{eq:poisson-family} was recognised by \citet{jaynes1968} as a scale-invariant likelihood in the context of Poisson processes.

The following thermometric models belong to the family in Eq.~\eqref{eq:poisson-family}. 
If we have a noninteracting gas of $n$ spin-$1/2$ particles, the probability that $x$ of them are found in the excited state is\footnote{
In denoting likelihood functions, we shall omit the dependency on dimensionless parameters which, as $n$ here, hold no relevance from the point of view of scale invariance.  
} \cite{jahnke2011, rubio2020, mehboudi2021},
\begin{equation}
    h\hspace{-0.1em}\left(x, \frac{y}{\theta}\right)
    = \binom{n}{x}\frac{\mathrm{exp}(-x y/\theta)}{Z(y/\theta)},
\end{equation}
where $y = \hbar \omega / k_B$, $\hbar$ is the reduced Planck's constant, $\omega$ is an angular frequency, and $Z(y/\theta) = [1 +  \mathrm{exp}(- y/\theta)]^n$.
Similarly, the probability that, for a quantum harmonic oscillator in thermal equilibrium, the dimensionless position coordinate $x$ lies between $x$ and $x + dx$ is written as \citep{pathria2011, rubio2020}
\begin{equation}
h\left(x, \frac{y}{\theta}\right) dx =
     \frac{\mathrm{exp}\lbrace - x^2/[2\,\sigma(y/\theta)^2] \rbrace }{\sqrt{2 \pi \sigma(y/\theta)^2}} dx,
\end{equation}
where $y=\hbar \omega / (2 k_B)$ and $2 \sigma(y/\theta)^2 = \mathrm{coth}(y/\theta)$.
Analogous considerations apply to the likelihood model for qubit thermometry in Ref.~\cite{boeyens2021noninformative}.

The class of models \eqref{eq:poisson-family} can be generalised by considering a collection of dimensionless measurands $\boldsymbol{x} = (x_1, \dots, x_{n})$, so that
\begin{equation}
p(\boldsymbol{x}| \theta,  y) = h\hspace{-0.1em}\left(\boldsymbol{x}, \frac{y}{\theta} \right).
\label{eq:collisional}
\end{equation}
Collisional thermometry, as formulated in Ref.~\cite{alves2021}, falls under this category. 
This is because the probability that (locally) measuring a block of $n$ ancillas generates a particular string of zeros and ones---i.e., $x_i = 0$ or $1$ for the $i$-th ancilla---has the form of Eq.~\eqref{eq:collisional}, with $y = \hbar \Omega/k_B$. 
Here, $\Omega$ is the resonant frequency associated with both the ancillas and an auxiliary system which connects the former with the system whose temperature is to be found; see Ref.~\cite{alves2021} for details. 

Now imagine that we further drop assumption (ii). 
Instead, let $x$ be a dimensioned measurand which is scaled by $T$, and assume that no other dimensioned parameters are involved. 
The condition to construct the likelihood function reads, in this case, $p(x|\theta)\,dx = p(x'|\theta')\,dx' = p(\gamma x|\gamma \theta) \gamma dx$. This leads to the functional equation $p(x|\theta) = \gamma p(\gamma x|\gamma \theta)$, whose solution is
\begin{equation}
p(x| \theta) =  h\hspace{-0.1em}\left(\frac{x}{\theta} \right) \frac{1}{\theta},
\label{eq:model-exp}
\end{equation}
for some function $h$.
Eq.~\eqref{eq:model-exp} formalises the notion of scale parameter traditionally employed in statistics \cite{mood1974}. 
Yet, we have seen that Eq.~\eqref{eq:model-exp} is not the only probability type leading to scale estimation problems. 

Thermal states with a continuous energy spectrum can be expressed as in Eq.~\eqref{eq:model-exp}.
Indeed, the probability that the energy $E$ given in units of temperature as $x = E/k_B$ lies between $x$ and $x+dx$ can be written as \cite{prosper1993,rubio2020},
\begin{align}
    p(x|\theta)\,dx &= f\hspace{-0.1em}\left(\frac{x}{\theta}\right)\left[\int_0^{\infty} d\breve{x}\,f\hspace{-0.1em}\left(\frac{\breve{x}}{\theta}\right)\right]^{-1} dx
    \nonumber \\
     &= f\hspace{-0.1em}\left(\frac{x}{\theta}\right)\left[\theta \int_0^{\infty} dt\,f(t)\right]^{-1} dx
    \nonumber \\
    &= h\hspace{-0.1em}\left(\frac{x}{\theta}\right) \frac{dx}{\theta},
    \nonumber
\end{align}
where $f$ is an arbitrary function, 
\begin{equation}
h\hspace{-0.1em}\left(\frac{x}{\theta}\right) = 
f\hspace{-0.1em}\left(\frac{x}{\theta}\right)\left[ \int_0^{\infty} dt\,f(t)\right]^{-1},
\end{equation}
and we have employed the change of variables $t = \breve{x}/\theta$.

While seemingly different, all these models are invariant under transformations $\theta \mapsto \theta' = \gamma \theta$ and either $y \mapsto y' = \gamma y$ or $x \mapsto x' = \gamma x$, and so temperature can here be interpreted as a scale parameter [Sec.~\ref{sec:formulation}]. 
Therefore, even when the operator-based results in Sec.~\ref{sec:main_result} do not apply for fixed likelihoods, temperature estimation based on either of these probability functions should arguably employ the formalism of global quantum thermometry in Ref.~\cite{rubio2020}, since the latter is a probability-based version of scale estimation theory.

\subsection{Quantum estimation of a lifetime}\label{sec:lifetime}

Now we turn to a different example illustrating the potential of quantum scale estimation beyond thermometry. 
Consider a two-level atom initially prepared in a superposition $\ket{\psi} = \sqrt{1-a} \ket{g} + \sqrt{a} \ket{e}$, where $\ket{g}$ and $\ket{e}$ denote its ground and excited states, respectively.
Due to the spontaneous emission of a photon, the excited state will decay to the ground state, a process whose statistics can be described by the density matrix \cite[Ch.~4]{2009barnett}  
\begin{align}
    \rho_t (\tau) =& \left[ 1 - a\,\eta_{t}(\tau) \right] \ketbra{g} +
    a\,\eta_{t}(\tau) \ketbra{e} 
    \nonumber \\
    & + 
    \left[a (1 - a)\,\eta_{t}(\tau)\right]^{\frac{1}{2}} (\ketbra{g}{e}+\ketbra{e}{g}).
    \label{eq:decay-state}
\end{align}
Here, $\eta_{t}(\tau) \coloneqq \exp(-t/\tau)$, $t$ is the elapsed time, and $\tau$ is the lifetime.
That is, the decay takes place at a rate $1/\tau$.

Imagine that the lifetime is unknown, i.e., $\Theta = \tau$, and that we wish to estimate it by observing whether or not a photon is emitted. 
Following Ref.~\cite[Ch.~4]{2009barnett}, this is captured by a POM with elements $M_{t,\tau_0}^Y = \left[1 - \eta_t(\tau_0) \right] \ketbra{e}$ (`Yes') and $M_{t,\tau_0}^N = \ketbra{g} + \eta_t(\tau_0) \ketbra{e}$ (`No'). 
To use this POM, we need an initial `hint', denoted by $\tau_o$, at the true lifetime $\tau$ \cite{rafal2015, rubio2020}.  
By combining such a physically-motivated measurement scheme with the state in Eq.~\eqref{eq:decay-state} via the Born rule, one arrives at the likelihood model $p(Y|\tau, \tau_0, t) = a \eta_t(\tau) [1 - \eta_t (\tau_0)]$ and $p(N|\tau, \tau_0, t) = 1 - p(Y|\tau, \tau_0, t)$. 
We then see that this is a scale estimation problem, since $p(Y|\tau, \tau_0, t) \mapsto p(Y|\theta, \boldsymbol{y}) = h_Y (\boldsymbol{y}/\theta)$, with $\boldsymbol{y} = (\tau_0, t)$ and $\theta$ playing the role of a hypothesis about the true value of $\tau$. 
As such, if the photon is detected, the optimal lifetime estimate is given by
\begin{align}
    \tilde{\vartheta}_{\boldsymbol{y}}^Y =
    \theta_u \exp\left[\int d\theta\, p(\theta|Y, \boldsymbol{y})\log\left(\frac{\theta}{\theta_u}\right)\right],
    \label{eq:decay-est}
\end{align}
where $p(\theta | Y, \boldsymbol{y}) \propto p(\theta)\, p(Y|\theta,\boldsymbol{y})$.
Otherwise, it is given by $\tilde{\vartheta}_{\boldsymbol{y}}^N$, with expression analogous to that in Eq.~\eqref{eq:decay-est}. 
Furthermore, the overall uncertainty is calculated as
\begin{align}
    \bar{\epsilon}_{\boldsymbol{y}, \mathrm{mle}} = & 
    \int d\theta\,p(Y,\theta|\boldsymbol{y}) \,\mathrm{log}^2\left(\frac{\tilde{\theta}^Y_{\boldsymbol{y}}}{\theta} \right)
     \nonumber \\
    & + \int d\theta\,p(N,\theta|\boldsymbol{y}) \,\mathrm{log}^2\left(\frac{\tilde{\theta}^N_{\boldsymbol{y}}}{\theta} \right).
    \label{eq:decay-mle}
\end{align}

To assess the quality of this strategy, Eq.~\eqref{eq:decay-mle} must be compared against two benchmarks: the uncertainty prior to performing the measurement, which we may denote as $\bar{\epsilon}_p$, and the fundamental limit to the precision associated with the state $\rho_t (\tau) $ in Eq.~\eqref{eq:decay-state}. 
The latter can be found by evaluating the quantum minimum $\bar{\epsilon}_{\boldsymbol{y}, \mathrm{min}}$ in Eq.~\eqref{eq:quantum_minimum} using $\rho_t (\tau) \mapsto \rho_y (\theta)$, with $y = t$.
As for the prior uncertainty, this is given as \cite{rubio2020}
\begin{equation}
    \bar{\epsilon}_p = \int d\theta\, p(\theta)\,\mathrm{log}^2\left(\frac{\theta}{\tilde{\vartheta}_p}\right),
    \label{eq:opt-prior-error}
\end{equation}
where
\begin{equation}
    \tilde{\vartheta}_p = \theta_u\,\mathrm{exp}\left[\int\,d\theta\, p(\theta)\,\mathrm{log}\left(\frac{\theta}{\theta_u}\right)\right]
    \label{eq:opt-est-prior}
\end{equation}
is the optimal prior estimate.

\begin{figure}[t]
\centering
\includegraphics[trim={0cm 0cm 0cm 0cm},clip,width=\linewidth]{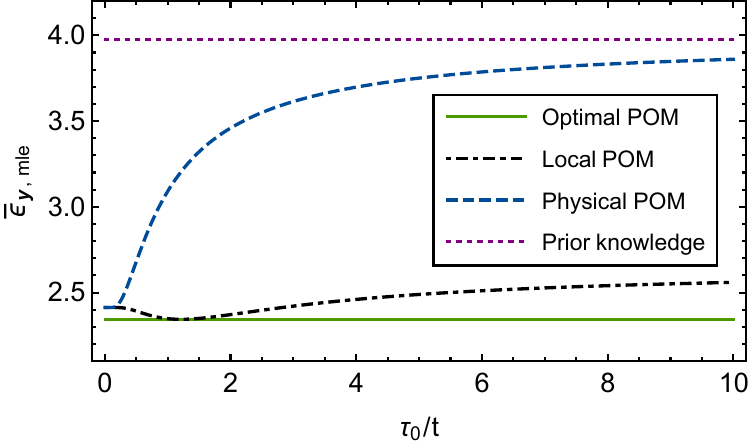}
	\caption{Quantum estimation of the lifetime of an atomic state. 
	The mean logarithmic errors of three measurement schemes are compared: one based on the optimal POM as predicted by quantum scale estimation (solid green), one using a physical POM representing whether or not a photon is detected (dashed blue), and one based on the POM predicted by \emph{local} quantum estimation theory \cite{paris2009, rafal2015}, hence referred to as `local POM' (dash-dotted black). 
    The latter two require an initial hint $\tau_0$ at the true value of the unknown lifetime \cite{rafal2015, rubio2020}, and this motivates examining the error as a function the ratio $\tau_0/t$, where $t$ is the time at which the measurement is performed. 
	The dotted purple line indicates the amount of knowledge available prior to performing the measurement. 
	As can be seen, the physical POM is informative in the sense that its associated error is below the prior uncertainty for the range of $\tau_0/t$, and it even becomes close to the fundamental limit when $\tau_0/t \ll 1$. 
	Yet, this is no longer the case when $\tau_0/t \gg 1$, implying that it is \emph{not} optimal in general.
    The local POM leads to a substantially smaller uncertainty in this regime, but it is still above the fundamental limit.
	See Sec.~\ref{sec:lifetime} for the details of this calculation.}
\label{fig:lifetime-est}
\end{figure}

For the sake of example, suppose we prepare the atom such that $a = 0.9$. 
Furthermore, consider the prior lifetime range $\theta/t \in [0.01, 10]$, so that the maximum-ignorance prior probability reads $p(\theta) = 0.145/\theta$. 
Given the dependence of the chosen POM on the initial hint $\tau_0$, it is natural to examine how the error changes as the ratio $\tau_0/t$ takes different values within the prior range. 
We shall perform this calculation numerically. 
The evaluation of Eqs.~\eqref{eq:opt-prior-error}, \eqref{eq:decay-mle} and \eqref{eq:quantum_minimum} lead, respectively, to the dotted purple curve in Fig.~\ref{fig:lifetime-est} for the prior uncertainty, the dashed blue line for the error associated with the physical POM, and the solid green line giving the fundamental limit to the precision for the atomic state in Eq.~\eqref{eq:decay-state}.
Note that both the prior uncertainty and the fundamental limit are independent of $\tau_0$, which is only relevant for the physical measurement, and so they appear as horizontal lines. 

Fig.~\ref{fig:lifetime-est} shows that the physical POM is informative to some extent; indeed, the prior uncertainty (dotted purple) is above its error (dashed blue) for the range of $\tau_0/t$.
However, we also see that the scheme extracts more information when $\tau_0/t \ll 1$ than when $\tau_0/t \gg 1$.
That is, it is easier to determine the lifetime in a regime where the decay is likely to have already happened, in consistency with our intuition. 
At the same time, the error for the physical POM becomes close to the fundamental limit (solid green) when $\tau_0/t \ll 1$, implying that the physical POM is almost-optimal in this regime, but this ceases to be the case as $\tau_0/t$ grows.
We thus conclude that the physical POM, while useful, is generally sub-optimal. 

We have seen that the framework in this work allows us to further determine which POM reaches the fundamental limit. 
Specifically, by calculating the eigenstates of the operator $\mathcal{S}_{y}$. 
Solving the Lyaponuv equation \eqref{eq:lyaponuv_eq} with the numerical values of this example leads to $\mathcal{S}_{y} = - 1.887 \ketbra{g} + 0.967 \ketbra{e} + 0.271 (\ketbra{g}{e} + \ketbra{e}{g})$.
Therefore, the optimal POM reaching the solid green line in Fig.~\ref{fig:lifetime-est} is given by $\mathcal{M}_y^{+} = \ketbra{\psi_{+}}$ and $\mathcal{M}_y^{-} = \ketbra{\psi_{-}}$, where $\ket{\psi_{+}} = 0.094 \ket{g} + 0.996 \ket{e}$ and $\ket{\psi_{-}} = 0.996 \ket{g} - 0.094 \ket{e}$. 
This demonstrates that quantum scale estimation can inform the design of optimal metrology schemes whenever the quantity to be estimated---in this case, a lifetime---is a scale parameter.

Finally, it is natural to ask whether the fundamental limit in this section can be reached via a POM given by the more familiar framework of \emph{local} quantum estimation theory. 
Specifically, by choosing the POM elements $M_{t,\tau_0}^i = |\lambda_{t,\tau_0}^i\rangle \langle \lambda_{t,\tau_0}^i|$, $i= 1, 2$, where $|\lambda_{t,\tau_0}^i\rangle$ is the $i$-th eigenstate of the symmetric logarithmic derivative $L_t(\tau_0)$ which is given by the Lyaponuv equation $L_t(\tau) \rho_t(\tau) + \rho_t(\tau) L_t(\tau) = 2\partial_\tau \rho_t(\tau)$ \cite{paris2009,rafal2015}. 
Using this local POM, a numerical calculation based on an expression analogous to that in Eq.~\eqref{eq:decay-mle} renders the black dash-dotted line in Fig.~\ref{fig:lifetime-est}.
As can be seen, the uncertainty associated with the local POM approaches the fundamental limit (solid green) as $\tau_0/t \ll 1$, and it actually reaches it when $\tau_0/t \sim 1$. 
However, it becomes sub-optimal for larger values of $\tau_0/t$, indicating that only quantum scale estimation can provide generally optimal POMs in the presence of finite prior knowledge.

\section{Discussion}\label{sec:discussion}

\subsection{The meaning of the operator $\mathcal{S}_{\boldsymbol{y}}$}\label{sec:Smeaning}

We have seen that, in quantum scale estimation, the operator $\mathcal{S}_{\boldsymbol{y}}$ is the key to find both the optimal quantum strategy and the associated minimum uncertainty.
It is thus desirable to identify its physical meaning. 

Motivated by the exponential map in Eq.~\eqref{eq:opt-est-final}, which connects the spectrum of $\mathcal{S}_{\boldsymbol{y}}$ with the optimal estimates for $\Theta$, consider the related construction
\begin{align}
    \hat{\Theta}_{\boldsymbol{y}} \coloneqq& \,\theta_u\,\mathrm{exp}(\mathcal{S}_{\boldsymbol{y}})
    \nonumber \\ 
     =& \,\theta_u \sum_{m=0}^\infty \frac{1}{m!} \left[\int ds\, \mathcal{P}_{\boldsymbol{y}}(s)\,s\right]^m
    \nonumber \\
    =& \,\theta_u \sum_{m=0}^\infty \frac{1}{m!} \int ds\,\mathcal{P}_{\boldsymbol{y}}(s)\,s^m
    \nonumber \\
    =& \,\theta_u \int ds\,\mathcal{P}_{\boldsymbol{y}}(s)\, \mathrm{exp}(s)
    \nonumber \\
    =& \int ds\,\mathcal{M}_{\boldsymbol{y}}(s)\,\tilde{\vartheta}_{\boldsymbol{y}}(s),
    \label{eq:new_observable}
\end{align}
where $\lbrace \tilde{\vartheta}_{\boldsymbol{y}}(s), \mathcal{M}_{\boldsymbol{y}}(s)\rbrace$ denotes the estimation strategy identified as optimal in Sec.~\ref{sec:opt-straegy}. 
We then recognise $\hat{\Theta}_{\boldsymbol{y}}$ as the optimal operator-valued estimator for the scale parameter $\Theta$, and $\mathcal{S}_{\boldsymbol{y}}$ is simply its log-transformed version $\mathcal{S}_{\boldsymbol{y}} = \mathrm{log}(\hat{\Theta}_{\boldsymbol{y}}/\theta_u)$.

\subsection{A quantum observable for scale parameters}\label{sec:obs}

The operator-valued estimator $\hat{\Theta}_{\boldsymbol{y}}$ plays, in addition, a more suggestive role.
Let an ideal measurement process be represented by the POM $\mathcal{P}_{\boldsymbol{y}}(s)$.
The operator $\mathcal{S}_{\boldsymbol{y}}$ is, in that case, a valid quantum observable \cite{personick1971,holevo2011}. 
Since one can construct new observables by taking functions over existing ones, the fact that $\mathcal{S}_{\boldsymbol{y}}$ is an observable implies that so is $\hat{\Theta}_{\boldsymbol{y}}$. 
Recalling that the eigenvalues of $\hat{\Theta}_{\boldsymbol{y}}$ are the values that can be experimentally assigned to the unknown parameter $\Theta$---indeed, they are estimates---it is then reasonable to treat $\hat{\Theta}_{\boldsymbol{y}}$ as a quantum observable for scale parameters.

This interpretation is in line with standard practice in quantum estimation theory \cite{helstrom1976,holevo2011}. 
Nevertheless, the physical implications of estimation-theoretic observables are not always fully appreciated. Recall that $\mathcal{S}_{\boldsymbol{y}}$---and thus $\hat{\Theta}_{\boldsymbol{y}}$---depends on: (i) the initial state and the parameter encoding, both captured by $\rho_{\boldsymbol{y}}(\theta)$; (ii) the prior information as represented by $p(\theta)$; and (iii) the fact that scale uncertainties are quantified using the mean logarithmic errors.
Then, the scale observable $\hat{\Theta}_{\boldsymbol{y}}$ serves to model scenarios where what we can observe may depend not only on the preparation of the experimental arrangement, as indicated by $\rho_{\boldsymbol{y}}(\theta)$, but also on what information is either available to us \emph{a priori} or simply deemed logically possible, which is what both $p(\theta)$ and $\bar{\epsilon}_{\boldsymbol{y},\mathrm{mle}}$ allow to quantify in a precise manner. 
The logarithmic uncertainty $\bar{\epsilon}_{\boldsymbol{y},\mathrm{mle}}$, in particular, is a type of noise-to-signal ratio mimicking the physics of scale parameters \cite{rubio2020}. 

An appealing aspect of observables such as $\hat{\Theta}_{\boldsymbol{y}}$ is that a well-defined procedure for their construction exists. 
Namely, the optimisation of error functionals such as $\bar{\epsilon}_{\boldsymbol{y},\mathrm{mle}}$. 
One could then take the notion of error functional as a primitive, search for the form of the deviation function which best suits the type of parameter of interest, find the estimation strategy which optimises such functional, and use it construct an observable for the aforementioned parameter. This procedure could in principle be applied to any quantity in physics, provided that one can frame it within an estimation problem. 

\subsection{Fundamental error bounds}\label{sec:fund-bounds}

We now focus on the fundamental minimum in Eq.~\eqref{eq:quantum_minimum}. 
While knowledge of such a minimum renders a discussion of bounds on $\bar{\epsilon}_{\boldsymbol{y},\mathrm{mle}}$ superfluous, rephrasing Eq.~\eqref{eq:quantum_minimum} as an inequality statement is useful to quantify the relative performance of measurements which are feasible in practice but different from the optimal POM in Eq.~\eqref{eq:opt-pom-final}. 
For example, as illustrated in Sec.~\ref{sec:lifetime} for the estimation of an atomic lifetime. 
More generally, consider the chain of inequalities
\begin{equation}
    \bar{\epsilon}_{\boldsymbol{y},\mathrm{mle}} \geq   
    \min_{\tilde{\theta}_{\boldsymbol{y}}}
    \mathcal{E}(\tilde{\theta}_{\boldsymbol{y}}, M_{\boldsymbol{y}})
    \geq   
    \min_{M_{\boldsymbol{y}}, \tilde{\theta}_{\boldsymbol{y}}}
    \mathcal{E}(\tilde{\theta}_{\boldsymbol{y}}, M_{\boldsymbol{y}}),
    \label{eq:qse-inequality}
\end{equation}
where 
\begin{equation}
    \mathcal{E}[\tilde{\theta}_{\boldsymbol{y}}(x), M_{\boldsymbol{y}}(x)] \coloneqq \mathrm{Tr}\left\lbrace\int dx\,M_{\boldsymbol{y}}(x)\,  W_{\boldsymbol{y}}[\tilde{\theta}_{\boldsymbol{y}}(x)]\right\rbrace.
    \label{eq:compact-def}
\end{equation}

From Ref.~\cite{rubio2020} we know that 
\begin{equation}
    \min_{\tilde{\theta}_{\boldsymbol{y}}}
    \mathcal{E}(\tilde{\theta}_{\boldsymbol{y}}, M_{\boldsymbol{y}}) = \bar{\epsilon}_p - \mathcal{K}_{\boldsymbol{y}},
    \label{eq:classical-min}
\end{equation}
where $\bar{\epsilon}_p$ is given in Eq.~\eqref{eq:opt-prior-error}, 
\begin{equation}
    \mathcal{K}_{\boldsymbol{y}} = \int dx\,p(x|\boldsymbol{y})\,\mathrm{log}^2\left[\frac{\tilde{\vartheta}_{\boldsymbol{y}}(x)}{\tilde{\vartheta}_p}\right]
\end{equation}
represents the information gained via the (not necessarily optimal) measurement $M_{\boldsymbol{y}}(x)$, $\tilde{\vartheta}_p$ and $\tilde{\vartheta}_{\boldsymbol{y}}(x)$ are given in Eqs.~\eqref{eq:opt-est-prior} and Eq.~\eqref{eq:optest-probability}, respectively.

Similarly, the minimum uncertainty found in this work, Eq.~\eqref{eq:quantum_minimum}, can be expressed as
\begin{align}
    \bar{\epsilon}_{\boldsymbol{y}, \mathrm{min}} &=
    \int d\theta\,p(\theta)\log^2\left(\frac{\theta}{\theta_u}\right) - \mathrm{Tr}(\varrho_{\boldsymbol{y}, 0} \mathcal{S}_{\boldsymbol{y}}^2)
    \nonumber \\
    &=\int d\theta\,p(\theta)\log^2\left(\frac{\theta}{\theta_u}\right) - \mathrm{Tr}\left[\varrho_{\boldsymbol{y}, 0} \log^2\left(\frac{\hat{\Theta}_{\boldsymbol{y}}}{\theta_u}\right)\right]
    \nonumber \\
    &= \bar{\epsilon}_p - \mathrm{Tr}\left[\varrho_{\boldsymbol{y}, 0} \log^2\left(\frac{\hat{\Theta}_{\boldsymbol{y}}}{\tilde{\vartheta}_p}\right)\right] \coloneqq \bar{\epsilon}_p - \mathcal{J}_{\boldsymbol{y}};
\end{align}
this stems from the operator-valued estimator $\hat{\Theta}_{\boldsymbol{y}}$ in Eq.~\eqref{eq:new_observable} plus choosing, without loss of generality, $\theta_u = \tilde{\vartheta}_p$.\footnote{
Note that choosing $\theta_u = \tilde{\vartheta}_p$ for the expression of the minimum is consistent with the definition of $\tilde{\vartheta}_p$. 
Indeed, inserting $\theta_u = \tilde{\vartheta}_p$ into Eq.~\eqref{eq:opt-est-prior} gives $0 = \int d\theta\,p(\theta)\,\mathrm{log}(\theta/\tilde{\vartheta}_p) = -\mathrm{log}(\tilde{\vartheta}_p/\breve{\theta}_u) + \int d\theta\,p(\theta)\,\mathrm{log}(\theta/\breve{\theta}_u)$, with arbitrary $\breve{\theta}_u$, from where Eq.~\eqref{eq:opt-est-prior} reemerges. 
}
Hence, 
\begin{equation}
    \min_{M_{\boldsymbol{y}},\,\tilde{\theta}_{\boldsymbol{y}}} 
    \mathcal{E}(\tilde{\theta}_{\boldsymbol{y}}, M_{\boldsymbol{y}}) = \bar{\epsilon}_p - \mathcal{J}_{\boldsymbol{y}}.
    \label{eq:quantum-min}
\end{equation}

Given Eqs.~\eqref{eq:classical-min} and \eqref{eq:quantum-min}, the chain of inequalities in Eq.~\eqref{eq:qse-inequality} can finally be written as
\begin{equation}
    \bar{\epsilon}_{\mathrm{mle}} \geq \bar{\epsilon}_p - \mathcal{K}_{\boldsymbol{y}} \geq \bar{\epsilon}_p - \mathcal{J}_{\boldsymbol{y}}.
    \label{eq:qse-inequality-form-final}
\end{equation}
The first inequality is saturated by processing the outcome $x$ with the optimal estimator $\tilde{\vartheta}_{\boldsymbol{y}}(x)$, and in principle there is no reason to choose any other estimator.\footnote{
Admittedly, there may be cases where the numerical computation of $\tilde{\vartheta}_{\boldsymbol{y}}(x)$ is challenging. 
If so, then it might be reasonable to work with a simpler estimator, provided that we also assess how close to the first bound in Eq.~\eqref{eq:qse-inequality-form-final} the associated uncertainty $\bar{\epsilon}_{\mathrm{mle}}$ is. 
}
The real advantage of Eq.~\eqref{eq:qse-inequality-form-final} lies in its second inequality: given $p(\theta)$, $\rho_{\boldsymbol{y}}(\theta)$ and a practical POM $M_{\boldsymbol{y}}(x)$, we can determine whether $M_{\boldsymbol{y}}(x)$ is optimal, almost-optimal \cite{berry2000} or manifestly sub-optimal by evaluating and comparing $\mathcal{K}_{\boldsymbol{y}}$ and $\mathcal{J}_{\boldsymbol{y}}$.
Importantly, such a procedure offers a computationally similar but more widely applicable alternative to the usual practice of comparing the classical Fisher information with its quantum counterpart.

It must be noted that this comparative study, while practical in scope, is generally theoretical in nature.
This is because $\bar{\epsilon}_{\boldsymbol{y},\mathrm{mle}}$ does not depend on a specific measurement outcome [Sec.~\ref{sec:formulation}], and so results based on it are to be understood as general statements about the class of schemes under analysis, rather than taking $\bar{\epsilon}_{\boldsymbol{y},\mathrm{mle}}$ as an error that must be measured in the laboratory. 
$\bar{\epsilon}_{\boldsymbol{y},\mathrm{mle}}$ can of course be `measured' if \emph{all} the probabilities involved in its construction happen to correspond to known distributions of relative frequencies; \citet{yan2018} examines this for errors of the same nature as $\bar{\epsilon}_{\boldsymbol{y},\mathrm{mle}}$. 
However, by no means needs the role of $\bar{\epsilon}_{\boldsymbol{y},\rm{mle}}$ be restricted in such a way, neither does an empirical $\bar{\epsilon}_{\boldsymbol{y},\rm{mle}}$ negate the necessity of reporting data processing errors via Eq.~\eqref{eq:err-exp} \cite{jaynes2003}.

\subsection{Multiparameter schemes}\label{sec:multiparameter}

This work has focused on single-parameter estimation, but realistic sensing schemes typically involve several unknown quantities \cite{szczykulska2016,demkowicz2020}. 
In our context, this type of scenario is illustrated, for example, by the family of probability models
\begin{equation}
    p(x|\boldsymbol{\theta}, \boldsymbol{y}) = \mathrm{Tr}[M_{\boldsymbol{y}}(x)\rho_{\boldsymbol{y}}(\boldsymbol{\theta})] = h\hspace{-0.1em}\left(x, \frac{y_1}{\theta_1}, \dots, \frac{y_d}{\theta_d}\right),
    \label{eq:multi-scale}
\end{equation}
where $\boldsymbol{\theta} = (\theta_1, \dots, \theta_d)$ denote hypotheses for the unknown parameters $\boldsymbol{\Theta} = (\Theta_1, \dots, \Theta_d)$, and each $y_i$ is scaled by $\Theta_i$.
While a general formulation is left for future work, it is interesting to explore---using Eq.~\eqref{eq:multi-scale}---a first approach to multiparameter scale estimation.
As in Sec.~\ref{sec:formulation}, we assume that knowledge of $\boldsymbol{y}$ does not inform the plausibility of the possible values for $\boldsymbol{\Theta}$.

First, one should arguably be sceptical about the existence of relationships between different parameters unless there is a reason (i.e., further information) to think otherwise. 
If we accept this viewpoint, any form of maximum ignorance in multiparameter estimation may be represented by a separable prior probability, i.e., $p(\boldsymbol{\theta}) = \prod_i p(\theta_i)$.

We further see that, as per Eq.~\eqref{eq:multi-scale}, maximum ignorance about each parameter amounts to imposing invariance under transformations $y_i \mapsto y_i' = \gamma_i y_i$ and $\theta_i \mapsto \theta_i' = \gamma_i\theta_i$ \cite{jaynes1968, toussaint2011}, with $0 <\gamma_i < \infty$.
Then, the condition $p(\boldsymbol{\theta})d\boldsymbol{\theta} = p(\boldsymbol{\theta}')d\boldsymbol{\theta}'$ leads to the functional equation $\prod_i p(\theta_i)=\prod_i \gamma_i\,p(\gamma_i\theta_i)$. 
Its solution, $p(\boldsymbol{\theta})\propto \prod_i (1/\theta_i)$ [Appendix~\ref{app:max-ig}], is simply the product of Jeffreys's prior \cite{jeffreys1961} for each parameter. 

Next, the transformation $\phi_i = \alpha_i \mathrm{log}(\theta_i/\theta_{u, i})$, with constant $\alpha_i$ and $\theta_{u,i}$, connects the estimation of a set of location parameters $\boldsymbol{\Phi} = (\Phi_1, \dots, \Phi_d)$, for which maximum ignorance is represented as $p(\boldsymbol{\phi}) = \prod_i p(\phi_i) \propto 1$, to our multiparameter scale estimation problem, in the sense that $p(\boldsymbol{\phi})d\boldsymbol{\phi} = p(\boldsymbol{\theta})d\boldsymbol{\theta}$ implies $p(\boldsymbol{\phi}) \propto 1 \mapsto p(\boldsymbol{\theta}) \propto \prod_i (1/\theta_i)$. 
Then, given a figure of merit for $\boldsymbol{\Phi}$, we may construct the figure of merit for $\boldsymbol{\Theta}$ by transforming the former. 

For a single location parameter $\phi_i$, the correct deviation function is the translation-invariant distance $|\tilde{\phi_i}(x)-\phi_i|^{k_i}$ \cite{jaynes2003, rubio2020}.  
For multiple location parameters, current practice considers the sum of distances $\mathcal{D}[\boldsymbol{\tilde{\phi}}(x),\boldsymbol{\phi}]=\sum_{i=1}^d |\tilde{\phi_i}(x)-\phi_i|^{k_i}/d$ \cite{humphreys2013,proctor2017networked}, where the $1/d$ factor indicates that all of them are taken to be equally important \cite{proctor2017networked,proctor2018}.
By transforming $\mathcal{D}(\boldsymbol{\tilde{\phi}},\boldsymbol{\phi})$ we thus find 
\begin{equation}
    \mathcal{D}(\boldsymbol{\tilde{\phi}},\boldsymbol{\phi}) 
    \mapsto
    \mathcal{D}(\boldsymbol{\tilde{\theta}},\boldsymbol{\theta})=\frac{1}{d}\sum_i \Bigg\vert \alpha_i \log{\left(\frac{\tilde{\theta}_i}{\theta_i}\right)}\Bigg\vert^{k_i},
    \label{eq:deviation-multiparameter}
\end{equation}
which is scale invariant. 
To recover a sum of noise-to-signal ratios in the limit of local prior knowledge, we take $\alpha_i = 1$ and $k_i = 2$, for all $i$ \cite{rubio2020}.

Having found a deviation function for several scale parameters, we can now use it to upgrade the notion of mean logarithmic error as
\begin{equation}
    \bar{\epsilon}_{\boldsymbol{y}, \mathrm{mle}} = \frac{1}{d}\sum_i \bar{\epsilon}_{\boldsymbol{y}, \mathrm{mle},i},
\end{equation}
where
\begin{equation}
    \bar{\epsilon}_{\boldsymbol{y}, \mathrm{mle},i} = \int dx\, d\boldsymbol{\theta}\, p(x,\boldsymbol{\theta}|\boldsymbol{y}) \log^2\left[\frac{\tilde{\theta}_{\boldsymbol{y},i}(x)}{\theta_i}\right]
    \label{eq:each-mle}
\end{equation}
and $p(x,\boldsymbol{\theta}|\boldsymbol{y}) = p(\boldsymbol{\theta}) \mathrm{Tr}[M_{\boldsymbol{y}}(x)\rho_{\boldsymbol{y}}(\boldsymbol{\theta})]$.
Note that this is consistent with the definition in Ref.~\cite{rubio2020}, as the new $\bar{\epsilon}_{\boldsymbol{y}, \mathrm{mle}}$ reduces to Eq.~\eqref{eq:th-mle} for $d=1$.

An immediate consequence of the new definition for $\bar{\epsilon}_{\boldsymbol{y},\mathrm{mle}}$ is that we can lower bound this error by finding the minimum value of each individual error $\bar{\epsilon}_{\boldsymbol{y},\mathrm{mle},i}$, since $\bar{\epsilon}_{\boldsymbol{y},\mathrm{mle},i} \geq 0$. 
Start by upgrading the definitions in Eqs.~\eqref{eq:quantum_moments} and \eqref{eq:Aop-aux} to
\begin{equation}
    \varrho_{\boldsymbol{y},k,i} = \int d\boldsymbol{\theta} \,p(\boldsymbol{\theta})\rho_{\boldsymbol{y}}(\boldsymbol{\theta}) \log^k\left(\frac{\theta_i}{\theta_{u,i}}\right)
    \label{eq:quantum_moments_multi}
\end{equation}
and
\begin{equation}
    \mathcal{A}_{\boldsymbol{y},k,i} = \int dx\,M_{\boldsymbol{y}}(x) \log^k\left[\frac{\tilde{\theta}_{\boldsymbol{y},i}(x)}{\theta_{u,i}}\right],
    \label{eq:Aop-aux_multi}
\end{equation}
respectively.
Note that $\varrho_{\boldsymbol{y},0,i} = \int d\boldsymbol{\theta} \,p(\boldsymbol{\theta})\rho_{\boldsymbol{y}}(\boldsymbol{\theta}) $ for all $i$.
Each error $\bar{\epsilon}_{\boldsymbol{y},\mathrm{mle},i}$ can then be rewritten as
\begin{align}
    \bar{\epsilon}_{\boldsymbol{y},\mathrm{mle},i} = & \int d\boldsymbol{\theta}\, p(\boldsymbol{\theta})\log^2\left(\frac{\theta_i}{\theta_{u,i}}\right) 
    \nonumber \\
    & + \mathrm{Tr}(\rho_{\boldsymbol{y},0,i} \mathcal{A}_{\boldsymbol{y},2,i} - 2 \rho_{\boldsymbol{y},1,i} \mathcal{A}_{\boldsymbol{y},1,i}).
    \label{eq:multi-quantumerror}
\end{align}
We see that minimising Eq.~\eqref{eq:multi-quantumerror} with respect to $\tilde{\theta}_{\boldsymbol{y},i}(x)$ and  $M_{\boldsymbol{y}}(x)$ is formally identical to the minimisation performed in Sec.~\ref{sec:min-derivation}. 
Consequently, $\bar{\epsilon}_{\boldsymbol{y}, \mathrm{mle}}$ is lower-bounded as
\begin{equation}
    \bar{\epsilon}_{\boldsymbol{y}, \mathrm{mle}} \geq \frac{1}{d} \sum_i \left[\int d\boldsymbol{\theta}\, p(\boldsymbol{\theta})\log^2\left(\frac{\theta_i}{\theta_{u,i}}\right) - \mathrm{Tr}(\rho_{\boldsymbol{y},0,i}\mathcal{S}_{\boldsymbol{y},i}^2)\right],
    \label{eq:QSEmultiparameter-bound}
\end{equation}
where each operator $\mathcal{S}_{\boldsymbol{y},i}$ solves the associated Lyapunov equation $\mathcal{S}_{\boldsymbol{y},i} \rho_{\boldsymbol{y},0,i} + \rho_{\boldsymbol{y},0,i} \mathcal{S}_{\boldsymbol{y},i} = 2 \rho_{\boldsymbol{y},1,i}$.

Eq.~\eqref{eq:QSEmultiparameter-bound} is a useful starting point to assess the overall uncertainty in the estimation of multiple scale parameters.  
An analogous result for the mean square error has in fact been proven informative in sensing networks and imaging \cite{jesus2020mar,lee2022}.
Yet, one can anticipate the inability to saturate the bound in Eq.~\eqref{eq:QSEmultiparameter-bound} whenever $[\mathcal{S}_{\boldsymbol{y},i},\mathcal{S}_{\boldsymbol{y},j}] \neq 0$ for $i\neq j$ \cite{jesus2020mar,demkowicz2020}. 
How is this incompatibility to be assessed and quantified is precisely the focus of current efforts in multiparameter quantum estimation \cite{sammy2016compatibility, carollo2019, albarelli2019, albarelli2019novB, mankei2020, demkowicz2020, sidhu2021}, as this is expected to shed new light on our understanding of the quantum-to-classical transition \cite{demkowicz2020}.
In that sense, bounds such as Eq.~\eqref{eq:QSEmultiparameter-bound} and that in Ref.~\cite{jesus2020mar} may help to frame the question of quantum compatibility such that its dependency on the prior information and the measure of uncertainty---which has been noted before \cite{alfredo2002}---is manifest.

\subsection{The role of invariance arguments}\label{sec:invariance}

We close on a conceptual note. 
In this work, probability theory is understood as an extension of the propositional calculus \cite{cox1946, paris1994, vanhorn2003}
capable of relating a set of propositions with the evidence that supports them \cite{ballentine2016}.
This approach, which we may call \emph{objective inference}, is one of the so-called Bayesian varieties of probability theory \cite{jaynes2003, ballentine2016}.
Within physics, this view gives rise to a notion of consistency such that two observers holding the same information must assign the same probability \cite{jaynes1968, jaynes2003}.
In turn, this leads to constraints that our probability functions must satisfy \cite{jaynes1968, toussaint2011}.

Invariance arguments \cite{kass1996, holevo2011, linden2014, larocca2022} offer one way of finding these constraints.
The estimation problem of interest is first enunciated as done at the beginning of Sec.~\ref{sec:formulation}. 
The set of propositions involved in this step constitute our \emph{state of information} \cite{jaynes2003, vanhorn2003}.
The task is then to identify which parameter transformations leave this state of information unaltered.
Once a relevant symmetry is identified---in our case, scale invariance---it can be used to construct, or at least constrain, the required probabilities.
This is the origin of the functional equations in Secs.~\ref{sec:formulation}, \ref{sec:other-laws} and \ref{sec:multiparameter}.

If a likelihood model is available, its functional form may reveal which parameter transformations are important.
These can then be used to calculate ignorance priors \cite{jaynes1968,toussaint2011}.
Such a procedure---which has been exploited in temperature estimation \cite{prosper1993,rubio2020} as well as in Sec.~\ref{sec:multiparameter}---is often interpreted as if invariance arguments could only be made with reference to a pre-existing likelihood model \cite[Ch.~10]{linden2014}. 
Yet, Sec.~\ref{sec:formulation} challenges this view by defining scale invariance \emph{before} a likelihood model---and even a quantum state---has been specified. 
In general, our state of information, and thus its associated symmetries, precedes the probability functions employed to represent it \cite[Ch.~12]{jaynes2003}.

States of information may sometimes satisfy no relevant symmetry. 
In that case, all potential reparametrisations---i.e., estimating some function $f(\Theta)$ rather than $\Theta$---could be seen as equally legitimate, motivating a demand for invariance under \emph{all} reparametrisations. 
That is,  a principle of indifference \cite{jaynes2003}.
This lack of symmetry might emerge, for instance, in temperature estimation with some non-equilibrium states; the generally-invariant version of global thermometry put forward by \citet{jorgensen2021bayesian} could provide a suitable framework there. 
Notwithstanding this, note that enforcing such a strong requirement on thermometric protocols which explicitly display simple symmetries---e.g., scale invariance in protocols based on Eq.~\eqref{eq:thermal_state}---would overconstrain our estimates, and this should be avoided for the sake of economy. 

It is also noted that the desideratum of invariance under all reparametrisations further leads to Jeffreys's general rule, upon which the prior probability $p(\theta)$ is taken as proportional to the square root of the Fisher information for a likelihood model \cite{1946jeffreys,kass1996}.
Using this rule has been suggested, for example, as a way to address `uninformed' thermometry \cite{boeyens2021noninformative}.
While superficially appealing, an immediate difficulty with this rule is that it violates the so-called likelihood principle \cite{jaynes2003, kass1996}.
Even if this is ignored, or bypassed by using the \emph{quantum} Fisher information---which only depends on the density operator \cite{rafal2015}---Jeffreys's general rule still has a strong and potentially unrealistic implication: that we know, \emph{a priori}, that the most likely values for the unknown parameter happen to be those for which the experimental setup is more sensitive.
This may be the case for exceedingly well-calibrated protocols, which \emph{are} realisable as per having abundant measurement data \cite{paris2009, rafal2015}.
Yet, for limited-data protocols \cite{jesus2018, jesus2019thesis, morelli2021}, the aforementioned approach of analysing which specific symmetries are at play \cite{jaynes1968, kass1996}, so that $p(\theta)$ is based on minimal assumptions, appears as a safer path to calculate truly uninformative priors.

The reasoning in this section can be challenged by switching to a different system of probability; the alternatives include subjective inference \cite{definetti1990, bernardo1994, ballentine2016}, pure frequency interpretations \cite{ballentine2016}, and the abstract random-variable formulation \cite{rosenthal2006}.
Yet, objective inference offers two advantages that together justify having advocated it in this work. 

\begin{table*}[t]
\setlength\extrarowheight{2pt}
\begin{tabular}{|l|c|c|c|} 
 \hline
 \textbf{Type of parameter} & phase & location & scale\\
 \hline
 \textbf{General support} & $0  \leq \theta < 2\pi$ & $-\infty < \theta < \infty$ & $0 < \theta < \infty$ \\
 \hline
 \textbf{Symmetry} & $\theta \mapsto \theta' = \theta + 2\gamma\pi,\,\gamma \in \mathbb{Z}$ & $\theta \mapsto \theta' = \theta + \gamma,\,\gamma \in \mathbb{R}$ & $\theta \mapsto \theta' = \gamma \theta,\,\gamma \in \mathbb{R}_{++}$ \\
 \hline
 \textbf{Maximum ignorance} & $p(\theta) = 1/2\pi$ & $p(\theta) \propto 1$ & $p(\theta) \propto 1/\theta$\\
 \hline
 \textbf{Deviation function} $\mathcal{D}(\tilde{\theta},\theta)$ & $4\,\mathrm{sin}^2[(\tilde{\theta}-\theta)/2]$ & $(\tilde{\theta}-\theta)^2$ & $\mathrm{log}^2(\tilde{\theta}/\theta)$ \\
 \hline
\end{tabular}
\caption{A basic prescription to develop quantum estimation theories for three of the most elementary notions in physics: phase, location and scale. 
The justification may be found, e.g., in Refs.~\cite{helstrom1976,holevo2011,rafal2015} for phases, Refs.~\cite{kass1996,jaynes2003} for locations, and Refs.~\cite{jaynes1968,prosper1993,rubio2020} for scales.}
\label{tab:basic-quantitites}
\end{table*}

First, not only does objective inference accommodate probabilities without a frequency interpretation, but it also recovers the usual correspondence with statistical frequencies whenever large numbers limits hold \cite{jaynes2003, rosenthal2006}.
Hence, use of this system of probability renders the widespread `Bayesian vs frequentist' divide unnecessary, both practically and conceptually.
Secondly, the logical---in the sense of taking propositions as a primitive---and impersonal language associated with this approach is arguably the most appropriate to search for mathematical principles that, on the basis of empirical data, account for natural phenomena and their laws---the aim of physics.

\section{Concluding remarks: phases, locations and scales}\label{sec:conclusions}

This work has addressed an important gap in quantum metrology.
Namely, current efforts gravitate around the estimation of phase and location parameters \cite{rafal2015,demkowicz2020}, using either periodic or square errors, but these are not appropriate when dealing with scale parameters \cite{rubio2020}.
In contrast, the framework of quantum scale estimation allows for the most precise measurement of scale parameters---as defined in Sec.~\ref{sec:formulation}---in a consistent manner, by following the optimal strategy in Eqs.~\eqref{eq:opt-pom-final} and \eqref{eq:opt-est-final}. 
Moreover, the minimum in Eq.~\eqref{eq:quantum_minimum} enables the search of fundamental limits to this precision, while the inequalities in Eq.~\eqref{eq:qse-inequality-form-final} give the means to assess how close to these limits can practical POMs be.

Since this framework has been formulated in a single-shot fashion, the next natural step is to consider multi-shot protocols and explore their asymptotic limits. 
Some results in this direction have been reported in Refs.~\cite{rubio2020,mehboudi2021} within the context of thermometry. 
In Ref.~\cite{rubio2020}, for example, the mean logarithmic error was shown to recover the quantum noise-to-signal ratio \cite{paris2009} in the limit of many measurement trials, establishing a connection with the classical Fisher information.
One would thus expect a link between the new framework and the \emph{quantum} Fisher information to exist. 
On a related note, it would be interesting to explore the potential connection between scale estimation as formulated here and the minimax approach \cite{gorecki2022}.

The most immediate use of the results in this work is to bridge the gap between the Bayesian approach to quantum thermometry currently under development \cite{rubio2020,alves2021,jorgensen2021bayesian,mehboudi2021,boeyens2021noninformative} and the full power of quantum estimation theory.
To that purpose, the study of thermal states in Sec.~\ref{sec:thermo-example}, together with the discussion of invariance principles in Sec.~\ref{sec:invariance}, provides a starting point. This could help, for instance, to clarify the link between Bayesian thermometry and other approaches to temperature estimation \cite{2015jarzyna, 2016pasquale, 2017pasquale, kiilerich2018, razavian2019} that may rely on non-equilibrium states.

Nonetheless, we have seen that quantum scale estimation is not restricted to thermometry. 
Namely, Sec.~\ref{sec:lifetime} has shown how the framework can be exploited to improve the measurement of the lifetime of an atomic state \cite{2009barnett}.
Furthermore, the multiparameter extension in Sec.~\ref{sec:multiparameter} may prove to be useful for the simultaneous estimation of multiple kinetic parameters---key in the biosciences \cite{baaske2014, subramanian2020, subramanian2021,eerqing2021}---when these are set to be measured using quantum states of light \cite{mpofu2021,mpofu2021exp}.

Beyond its applicability, quantum scale estimation offers a wider theoretical perspective. 
When the focus is mainly---and perhaps inadvertently---placed on square errors, one is tempted to believe that results based on $\mathcal{D}(\tilde{\theta},\theta) = (\tilde{\theta} - \theta)^2$ are applicable to any possible parameter of interest.
Although consideration of phases alleviates this to some extent, the approximation $\mathcal{D}(\tilde{\theta},\theta) = 4 \sin^2 [ (\tilde{\theta} - \theta)/2 ]\approx (\tilde{\theta} - \theta)^2$ \cite{rafal2015} encourages the idea that square errors may still suffice in a somehow wide range of cases. 
The need for relative errors, as is $\mathcal{D}(\tilde{\theta},\theta) = \log^2(\tilde{\theta}/ \theta)$ \cite{rubio2020,jorgensen2021bayesian}, manifestly reduces such range, and it suggests a more attractive viewpoint: that three of the most elementary concepts in physics---phase, location and scale---should each have a dedicated quantum estimation theory, in accordance with Tab.~\ref{tab:basic-quantitites}. 
%---------------------------------------------------------

\begin{acknowledgments}
The author gratefully thanks M. R. J\o rgensen, J. Glatthard, J. Dunningham, S. Moore, and D. Porras for insightful exchanges on the notions of scale parameter and quantum observable, and F. Cerisola, J. Anders, K. Burrows, L. A. Correa, S. Scali, G. Alves, J. Boeyens, S. Nimmrichter, and O. Kyriienko for helpful discussions and comments. 
The author also acknowledges support from the United Kingdom EPSRC (Grants No. EP/T002875/1 and EP/R045577/1).
\end{acknowledgments}

\appendix

\section{Minimisation of $\mathrm{Tr}(\varrho_{\boldsymbol{y}, 0} \mathcal{A}_{\boldsymbol{y},1}^2 - 2\varrho_{\boldsymbol{y}, 1} \mathcal{A}_{\boldsymbol{y},1})$ over $\mathcal{A}_{\boldsymbol{y},1}$}\label{app:minimisation_aux}

This appendix uses the calculus of variations in operator form to find
\begin{equation}
    \min_{\mathcal{A}_{\boldsymbol{y},1}} f(\mathcal{A}_{\boldsymbol{y},1}),
    \label{eq:minimum_app}
\end{equation}
where $f(\mathcal{A}_{\boldsymbol{y},1}) = \mathrm{Tr}(\varrho_{\boldsymbol{y}, 0} \mathcal{A}_{\boldsymbol{y},1}^2 - 2\varrho_{\boldsymbol{y}, 1} \mathcal{A}_{\boldsymbol{y},1})$.

The condition for a given $\mathcal{A}_{\boldsymbol{y},1}$ to give rise to an extremum is \cite{mathematics2004, personick1971}
\begin{equation}
    \frac{d f(\mathcal{A}_{\boldsymbol{y},1}+\alpha \Gamma)}{d\alpha}\bigg\rvert_{\alpha = 0} = 0,
    \label{eq:variation-condition}
\end{equation}
where $\Gamma$ denotes a Hermitian but otherwise arbitrary operator variation.
Given that the first variation of $f(\mathcal{A}_{\boldsymbol{y},1})$ reads 
\begin{equation}
    \frac{d f(\mathcal{A}_{\boldsymbol{y},1}+\alpha \Gamma)}{d\alpha}\bigg\rvert_{\alpha = 0} = \mathrm{Tr}[(\mathcal{A}_{\boldsymbol{y},1} \varrho_{\boldsymbol{y}, 0} + \varrho_{\boldsymbol{y}, 0} \mathcal{A}_{\boldsymbol{y},1} - 2 \varrho_{\boldsymbol{y}, 1})\Gamma],
    \label{eq:variation1}
\end{equation}
Eq.~\eqref{eq:variation-condition} amounts to impose
\begin{equation}
    \mathrm{Tr}[(\mathcal{A}_{\boldsymbol{y},1} \varrho_{\boldsymbol{y}, 0} + \varrho_{\boldsymbol{y}, 0} \mathcal{A}_{\boldsymbol{y},1} - 2 \varrho_{\boldsymbol{y}, 1})\Gamma] = 0
    \label{eq:condition-equation}
\end{equation}
for all $\Gamma$. This is satisfied when
\begin{equation}
    \mathcal{A}_{\boldsymbol{y},1} \varrho_{\boldsymbol{y}, 0} + \varrho_{\boldsymbol{y}, 0} \mathcal{A}_{\boldsymbol{y},1} = 2\varrho_{\boldsymbol{y}, 1},
\end{equation}
which is a Lyaponuv equation for $\mathcal{A}_{\boldsymbol{y},1}$.
For the sake of clarity, we may denote the operator $\mathcal{A}_{\boldsymbol{y},1}$ that is solution to such equation by $\mathcal{S}_{\boldsymbol{y}}$, i.e., 
\begin{equation}
    \mathcal{S}_{\boldsymbol{y}} \varrho_{\boldsymbol{y}, 0} + \varrho_{\boldsymbol{y}, 0} \mathcal{S}_{\boldsymbol{y}} = 2\varrho_{\boldsymbol{y}, 1}.
    \label{eq:condition_for_minimum}
\end{equation}

So far we have that $\mathcal{A}_{\boldsymbol{y},1} = \mathcal{S}$ makes the quantity $f(\mathcal{A}_{\boldsymbol{y},1})$ extremal. Given that
\begin{equation}
    \frac{d^2 f(\mathcal{A}_{\boldsymbol{y},1}+\alpha \Gamma)}{d\alpha^2}\bigg\rvert_{\alpha = 0} =2\,\mathrm{Tr}(\varrho_{\boldsymbol{y}, 0}\Gamma^2) > 0,
    \label{eq:proving-minimum}
\end{equation}
we can conclude that, in fact, $\mathcal{A}_{\boldsymbol{y},1} = \mathcal{S}_{\boldsymbol{y}}$ gives rise to a minimum. 
To see the validity of Eq.~\eqref{eq:proving-minimum}, first consider the eigendecomposition $\Gamma = \int d\gamma\,\mathcal{P}(\gamma)\,\gamma$. 
In that case, 
\begin{align}
    \mathrm{Tr}(\varrho_{\boldsymbol{y}, 0} \Gamma^2) & =
    \int d\gamma d\breve{\gamma}\,\mathrm{Tr}[\varrho_{\boldsymbol{y}, 0} \mathcal{P}(\gamma) \mathcal{P}(\breve{\gamma})] \,\gamma\,\breve{\gamma}
    \nonumber \\
    & = \int d\gamma d\breve{\gamma}\,\delta(\gamma - \breve{\gamma})\,\mathrm{Tr}[\mathcal{P}(\breve{\gamma}) \varrho_{\boldsymbol{y}, 0}] \,\gamma\,\breve{\gamma}
    \nonumber \\
    & = \int d\gamma\,\mathrm{Tr}[\mathcal{P}(\gamma)\varrho_{\boldsymbol{y}, 0}]\,\gamma^2
    \nonumber \\
    & = \int d\gamma\,p(\gamma | \boldsymbol{y})\,\gamma^2,
    \label{eq:aux-variations}
\end{align}
so that Eq.~\eqref{eq:proving-minimum} holds.
The last equality in Eq.~\eqref{eq:aux-variations} is found as in Eq.~\eqref{eq:opt-evidence}.

Finally, by inserting $\mathcal{A}_{\boldsymbol{y},1} = \mathcal{S}_{\boldsymbol{y}}$ into $f(\mathcal{A}_{\boldsymbol{y},1})$, and using the fact that $\mathcal{S}_{\boldsymbol{y}} \varrho_{\boldsymbol{y}, 0} + \varrho_{\boldsymbol{y}, 0} \mathcal{S}_{\boldsymbol{y}} = 2\varrho_{\boldsymbol{y}, 1}$ implies $\mathrm{Tr}(\varrho_{\boldsymbol{y}, 0} \mathcal{S}_{\boldsymbol{y}}^2) = \mathrm{Tr}(\varrho_{\boldsymbol{y}, 1} \mathcal{S}_{\boldsymbol{y}}) $, we find the minimum to be 
\begin{equation}
    \min_{\mathcal{A}_{\boldsymbol{y},1}} f(\mathcal{A}_{\boldsymbol{y},1}) = f(\mathcal{S}_{\boldsymbol{y}}) = - \mathrm{Tr}(\varrho_{\boldsymbol{y}, 0} \mathcal{S}_{\boldsymbol{y}}^2),
    \label{eq:min-appendix}
\end{equation}
as stated in the main text. 

\section{The fundamental equations of the optimal quantum strategy}\label{app:HH-path}

Upon adapting it to scale estimation, a fundamental result given by Holevo \cite{holevo1973,holevo1973b} and Helstrom \cite{helstrom1974,helstrom1976} is as follows: if $\mathcal{M}_{\boldsymbol{y}}(z)$ is the POM that minimises the overall uncertainty $\bar{\epsilon}_{\boldsymbol{y}, \mathrm{mle}}$---hence called optimal---and $\tilde{\vartheta}_{\boldsymbol{y}}(z)$ is the associated estimator, then there exists an operator
\begin{equation}
    \Upsilon_{\boldsymbol{y}} = \int dz\,\mathcal{M}_{\boldsymbol{y}}(z)\,W_{\boldsymbol{y}}[\tilde{\vartheta}_{\boldsymbol{y}}(z)]
    =\int dz\,W_{\boldsymbol{y}}[\tilde{\vartheta}_{\boldsymbol{y}}(z)]\,\mathcal{M}_{\boldsymbol{y}}(z)
    \label{eq:HH_eqs1}
\end{equation}
satisfying 
\begin{equation}
    W_{\boldsymbol{y}}[\tilde{\vartheta}_{\boldsymbol{y}}(z)] - \Upsilon_{\boldsymbol{y}} \geq 0 
    \label{eq:HH_eqs2}
\end{equation}
and leading to the minimum uncertainty $\bar{\epsilon}_{\boldsymbol{y},\mathrm{min}} = \mathrm{Tr}(\Upsilon_{\boldsymbol{y}})$.

Sec.~\ref{sec:opt-straegy} showed that the optimal strategy for quantum scale estimation is given by
\begin{equation}
    \mathcal{M}_{\boldsymbol{y}}(z) \mapsto \mathcal{M}_{\boldsymbol{y}}(s) = \mathcal{P}_{\boldsymbol{y}}(s)
    \label{eq:optimal_strategy}
\end{equation}
and 
\begin{equation}
    \tilde{\vartheta}(z) \mapsto \tilde{\vartheta}(s) = \theta_u \mathrm{exp}(s),
    \label{eq:optimal_strategy2}
\end{equation}
where $\mathcal{P}_{\boldsymbol{y}}(s)$ is the projector associated with the eigendecomposition $\mathcal{S}_{\boldsymbol{y}} = \int ds\,\mathcal{P}_{\boldsymbol{y}}(s)\,s$.
To see that this strategy does indeed verify Eq.~\eqref{eq:HH_eqs2}, we now follow the steps of the related proof for the mean square error in Ref.~\cite[Ch.~8]{helstrom1976}. 

By inserting Eqs.~\eqref{eq:optimal_strategy} and \eqref{eq:optimal_strategy2} into Eq.~\eqref{eq:aux_W}, we find that $W_{\boldsymbol{y}}[\tilde{\vartheta}_{\boldsymbol{y}}(s)] = W_{\boldsymbol{y}}[\theta_u \mathrm{exp}(s)] = \varrho_{\boldsymbol{y},2} + \varrho_{\boldsymbol{y}, 0} s^2 - 2 \varrho_{\boldsymbol{y}, 1} s$, and so
\begin{equation}
    \Upsilon_{\boldsymbol{y}} = \varrho_{\boldsymbol{y}, 2} + \mathcal{S}_{\boldsymbol{y}}^2\varrho_{\boldsymbol{y}, 0} - 2 \mathcal{S}_{\boldsymbol{y}} \varrho_{\boldsymbol{y}, 1} =\varrho_{\boldsymbol{y}, 2} + \varrho_{\boldsymbol{y}, 0} \mathcal{S}_{\boldsymbol{y}}^2 - 2\varrho_{\boldsymbol{y}, 1} \mathcal{S}_{\boldsymbol{y}}.
    \label{eq:two_forms_aux}
\end{equation}
Using Eq.~\eqref{eq:lyaponuv_eq} and the two forms of $\Upsilon_{\boldsymbol{y}}$ in Eq.~\eqref{eq:two_forms_aux}, we can write 
\begin{equation}
    W_{\boldsymbol{y}}[\tilde{\theta}_{\boldsymbol{y}}(s)]=\varrho_{\boldsymbol{y}, 2} + s\varrho_{\boldsymbol{y}, 0} s - \mathcal{S}_{\boldsymbol{y}}\varrho_{\boldsymbol{y}, 0} s - s \varrho_{\boldsymbol{y}, 0} \mathcal{S}_{\boldsymbol{y}}
\end{equation}
and
\begin{equation}
    \Upsilon_{\boldsymbol{y}} = \varrho_{\boldsymbol{y}, 2} - \mathcal{S}_{\boldsymbol{y}}\varrho_{\boldsymbol{y}, 0}\mathcal{S}_{\boldsymbol{y}},
\end{equation}
so that $ W_{\boldsymbol{y}}[\tilde{\theta}_{\boldsymbol{y}}(s)]-\Upsilon_{\boldsymbol{y}} = (\mathcal{S}_{\boldsymbol{y}}-s)\varrho_{\boldsymbol{y}, 0} (\mathcal{S}_{\boldsymbol{y}}-s)$. 
By using this and the fact that $\mathcal{S}_{\boldsymbol{y}}^\dagger = \mathcal{S}_{\boldsymbol{y}}$, we find that
\begin{align}
    \bra{v} \lbrace W[\tilde{\theta}_{\boldsymbol{y}}(s)]-\Upsilon_{\boldsymbol{y}} \rbrace \ket{v} &= \bra{v}(\mathcal{S}_{\boldsymbol{y}}-s)\varrho_{\boldsymbol{y}, 0} (\mathcal{S}_{\boldsymbol{y}}-s)\ket{v}
    \nonumber \\
    &= \bra{v'}\varrho_{\boldsymbol{y}, 0} \ket{v'} \geq 0,
    \label{eq:HH_proof_aux}
\end{align}
where $\ket{v'}=(\mathcal{S}_{\boldsymbol{y}}-s)\ket{v}$ and $\ket{v}$ is an arbitrary ket. Eq.~\eqref{eq:HH_proof_aux} is equivalent to Eq.~\eqref{eq:HH_eqs2}, as required. 
Moreover, note that $\mathrm{Tr}(\Upsilon_{\boldsymbol{y}}) = \mathrm{Tr}(\varrho_{\boldsymbol{y}, 2} - \mathcal{S}_{\boldsymbol{y}}\varrho_{\boldsymbol{y}, 0} \mathcal{S}_{\boldsymbol{y}})$ leads to the minimum mean square error in Eq.~\eqref{eq:quantum_minimum}.

\section{Alternative proof of Eqs.~\eqref{eq:quantum_minimum}, \eqref{eq:opt-pom-final} and \eqref{eq:opt-est-final}}\label{app:JJ-path}

This appendix recovers the results in Sec.~\ref{sec:main_result} using a different method---that in Ref.~\cite{jesus2020mar}. 
Start by minimising $\bar{\epsilon}_{\boldsymbol{y},\mathrm{mle}} = \mathcal{E}[\tilde{\theta}_{\boldsymbol{y}}(x),\,M_{\boldsymbol{y}}(x)]$ over the estimator $\tilde{\theta}_{\boldsymbol{y}}(x)$ while keeping the POM $M_{\boldsymbol{y}}(x)$ fixed. Here, $\mathcal{E}$ is defined as in Eq.~\eqref{eq:compact-def}.
As per Ref.~\cite{rubio2020}, this gives
\begin{equation}
    \min_{\tilde{\theta}_{\boldsymbol{y}}(x)} \mathcal{E}[\tilde{\theta}_{\boldsymbol{y}}(x),\,M_{\boldsymbol{y}}(x)] = A - B_{\boldsymbol{y}}[M_{\boldsymbol{y}}(x)],
    \label{eq:first_minimum_app}
\end{equation}
where
\begin{equation}
    A = \int d\theta\,p(\theta)\,\mathrm{log}^2\left(\frac{\theta}{\theta_u}\right)
    \label{eq:prior_term_app}
\end{equation}
is a measurement-independent quantity, the POM enters in $B[M_{\boldsymbol{y}}(x)]$ as
\begin{equation}
    B[M_{\boldsymbol{y}}(x)] = \int dx\,p(x|\boldsymbol{y})\,\mathrm{log}^2\left[\frac{\tilde{\vartheta}_{\boldsymbol{y}}(x)}{\theta_u}\right],
    \label{eq:min_past_paper}
\end{equation}
and $\tilde{\vartheta}_{\boldsymbol{y}}(x)$ is the optimal estimator given by 
\begin{equation}
    \tilde{\vartheta}_{\boldsymbol{y}}(x) = \theta_u \exp\left[\int d\theta\,p(\theta|x, \boldsymbol{y})\log\left(\frac{\theta}{\theta_u}\right)\right].
    \label{eq:optimal_est_app}
\end{equation}

Next, use the Born rule $p(x|\theta, \boldsymbol{y}) = \mathrm{Tr}[M_{\boldsymbol{y}}(x) \rho_{\boldsymbol{y}}(\theta)]$ to write $p(x|\boldsymbol{y}) = \mathrm{Tr}[M_{\boldsymbol{y}}(x) \varrho_{\boldsymbol{y}, 0}]$,
\begin{equation}
    \tilde{\vartheta}_{\boldsymbol{y}}(x) = \theta_u \exp\left\lbrace\frac{\mathrm{Tr}[M_{\boldsymbol{y}}(x)\varrho_{\boldsymbol{y}, 1} ]}{\mathrm{Tr}[M_{\boldsymbol{y}}(x)\varrho_{\boldsymbol{y}, 0} ]}\right\rbrace
    \label{eq:optimal_est_app2}
\end{equation}
and, as a consequence,
\begin{equation}
    B_{\boldsymbol{y}}[M_{\boldsymbol{y}}(x)] = \int dx\,\frac{\mathrm{Tr}[ M_{\boldsymbol{y}}(x)\varrho_{\boldsymbol{y}, 1}]^2}{\mathrm{Tr}[M_{\boldsymbol{y}}(x)\varrho_{\boldsymbol{y}, 0} ]},
    \label{eq:aux_app}
\end{equation}
where $\varrho_{\boldsymbol{y}, 0}$ and $\varrho_{\boldsymbol{y}, 1}$ have been defined in Sec.~\ref{sec:min-derivation}.

Eq.~\eqref{eq:aux_app} is formally equivalent to the starting point for the proof of the celebrated Braunstein-Caves inequality \cite{braunstein1994,genoni2008}, as well as of its Bayesian analogue \cite{jesus2020mar, jesus2019thesis}. 
Following the structure of such proofs, \emph{redefine} the operator $\varrho_{\boldsymbol{y}, 1}$ as $\varrho_{\boldsymbol{y}, 1} = (\mathcal{X}_{\boldsymbol{y}}\varrho_{\boldsymbol{y}, 0} + \varrho_{\boldsymbol{y}, 0}\mathcal{X}_{\boldsymbol{y}})/2$. 
At this point, the meaning of $\mathcal{X}_{\boldsymbol{y}}$ is unknown. 
However, to simplify the notation we may rename such $\mathcal{X}_{\boldsymbol{y}}$ as $\mathcal{S}_{\boldsymbol{y}}$ in anticipation to the label employed in the main text. 
If we use this to rewrite Eq.~\eqref{eq:aux_app}, then we can upper-bound $B_{\boldsymbol{y}}[M_{\boldsymbol{y}}(x)]$, for any $p(\theta)$ and $\rho_{\boldsymbol{y}}(\theta)$, as
\begin{align}
    B_{\boldsymbol{y}}[M_{\boldsymbol{y}}(x)] &= \int dx \left(\frac{\mathrm{Re}\lbrace \mathrm{Tr}[M_{\boldsymbol{y}}(x) \mathcal{S}_{\boldsymbol{y}}\rho_{\boldsymbol{y},0}]
    \rbrace}{\sqrt{\mathrm{Tr}[M_{\boldsymbol{y}}(x)\rho_{\boldsymbol{y},0}]}}\right)^2
    \nonumber \\
    &\leq \int dx\,\Bigg\rvert \frac{\mathrm{Tr}[M_{\boldsymbol{y}}(x) \mathcal{S}_{\boldsymbol{y}}\rho_{\boldsymbol{y},0}]}{\sqrt{\mathrm{Tr}[M_{\boldsymbol{y}}(x)\rho_{\boldsymbol{y},0}]}}\Bigg\lvert^2
    \nonumber \\
    & = \int dx\,\Bigg\rvert\mathrm{Tr}\left[\frac{\varrho_{\boldsymbol{y}, 0}^{\frac{1}{2}}M_{\boldsymbol{y}}(x)^{\frac{1}{2}}}{\sqrt{\mathrm{Tr}[M_{\boldsymbol{y}}(x)\rho_{\boldsymbol{y},0}]}} M_{\boldsymbol{y}}(x)^{\frac{1}{2}}\mathcal{S}_{\boldsymbol{y}}\varrho_{\boldsymbol{y}, 0}^{\frac{1}{2}}\right]\Bigg\lvert^2
    \nonumber \\
    & \leq \int dx\,\mathrm{Tr}[M_{\boldsymbol{y}}(x)\mathcal{S}_{\boldsymbol{y}}\varrho_{\boldsymbol{y}, 0} \mathcal{S}_{\boldsymbol{y}}] 
    \nonumber \\
    & = \mathrm{Tr}(\varrho_{\boldsymbol{y}, 0} \mathcal{S}_{\boldsymbol{y}}^2),
    \label{eq:caves_braunstein}
\end{align}
where we have used the Cauchy-Schwarz inequality \cite{helstrom1968multiparameter}
\begin{equation}
    |\mathrm{Tr}(Y^\dagger Z)|^2 \leq \mathrm{Tr}(Y^\dagger Y) \,\mathrm{Tr}(Z^\dagger Z),
\end{equation}
with
\begin{equation}
    Y = \frac{M_{\boldsymbol{y}}(x)^{\frac{1}{2}}\varrho_{\boldsymbol{y}, 0}^{\frac{1}{2}}}{\sqrt{\mathrm{Tr}[M_{\boldsymbol{y}}(x)\rho_{\boldsymbol{y},0}]}}, 
    \hspace{0.3cm}
    Z = M_{\boldsymbol{y}}(x)^{\frac{1}{2}}\mathcal{S}_{\boldsymbol{y}}\varrho_{\boldsymbol{y}, 0}^{\frac{1}{2}}.
\end{equation}
The combination of Eqs.~\eqref{eq:first_minimum_app}, \eqref{eq:prior_term_app} and \eqref{eq:caves_braunstein} thus leads to
\begin{equation}
    \bar{\mathrm{\epsilon}}_{\boldsymbol{y},\mathrm{mle}} \geq \int d\theta\,p(\theta)\,\mathrm{log}^2\left(\frac{\theta}{\theta_u}\right) - \mathrm{Tr}(\varrho_{\boldsymbol{y}, 0} \mathcal{S}_{\boldsymbol{y}}^2).
    \label{eq:inequality-app}
\end{equation}

A real-valued $\mathrm{Tr}[M_{\boldsymbol{y}}(x) \mathcal{S}_{\boldsymbol{y}}\varrho_{\boldsymbol{y}, 0}]$ saturates the first inequality in Eq.~\eqref{eq:caves_braunstein}. 
The Cauchy-Schwarz inequality is saturated when $Y \propto Z$ \cite{helstrom1968multiparameter}, which implies that
\begin{equation}
    \frac{M_{\boldsymbol{y}}(x)^{\frac{1}{2}}\varrho_{\boldsymbol{y}, 0}^{\frac{1}{2}}}{\mathrm{Tr}[M_{\boldsymbol{y}}(x)\varrho_{\boldsymbol{y}, 0}]} = \frac{M_{\boldsymbol{y}}(x)^{\frac{1}{2}}\mathcal{S}_{\boldsymbol{y}}\varrho_{\boldsymbol{y}, 0}^{\frac{1}{2}}}{\mathrm{Tr}[M_{\boldsymbol{y}}(x)\mathcal{S}_{\boldsymbol{y}}\varrho_{\boldsymbol{y}, 0}]}. 
\end{equation}
By choosing
\begin{equation}
    M_{\boldsymbol{y}}(x) \mapsto \mathcal{M}_{\boldsymbol{y}}(s) = \mathcal{P}_{\boldsymbol{y}}(s),
    \label{eq:opt_measurement_app}
\end{equation}
where $\mathcal{P}_{\boldsymbol{y}}(s)$ is the projector associated with $\mathcal{S}_{\boldsymbol{y}}$, both saturation conditions are fulfilled at once \cite{genoni2008, jesus2020mar}. 
Furthermore, inserting Eq.~\eqref{eq:opt_measurement_app} into Eq.~\eqref{eq:optimal_est_app2} allows us to write the optimal estimator as
\begin{equation}
    \tilde{\vartheta}_{\boldsymbol{y}}(s) = \theta_u \mathrm{exp}\left\lbrace\frac{\mathrm{Tr}[ \mathcal{P}_{\boldsymbol{y}}(s)\varrho_{\boldsymbol{y}, 1}]}{\mathrm{Tr}[ \mathcal{P}_{\boldsymbol{y}}(s)\varrho_{\boldsymbol{y}, 0}]}\right\rbrace = \theta_u\,\mathrm{exp}(s),
    \label{eq:opt_est_app_final}
\end{equation}
where the second inequality stems from Eq.~\eqref{eq:gqt-qse_equivalence}.

Given that the equality in Eq.~\eqref{eq:inequality-app} can be reached by using the measurement in Eq.~\eqref{eq:opt_measurement_app} and the estimator in Eq.~\eqref{eq:opt_est_app_final}, we identify it as the minimum associated with our original optimisation problem in Eq.~\eqref{eq:min_def}. 
As expected, it is identical to that in Eq.~\eqref{eq:quantum_minimum}, as are Eqs.~\eqref{eq:opt_measurement_app} and \eqref{eq:opt_est_app_final} to  Eqs.~\eqref{eq:opt-pom-final} and \eqref{eq:opt-est-final} in the main text, respectively. 

\section{A multiparameter prior density}\label{app:max-ig}

To have a notion of maximum ignorance in multiparameter scale estimation, Sec.~\ref{sec:multiparameter} identifies the constraint $\prod_i p(\theta_i) = \prod_i \gamma_i\,p(\gamma_i\theta_i)$. 
This functional equation may be solved as follows. 
Rewrite it as
\begin{equation}
    \gamma_1 p(\gamma_1 \theta_1) = p(\theta_1) \prod_{i\neq 1} \frac{p(\theta_i)}{\gamma_i\,p(\gamma_i\theta_i)}
    \label{eq:functional_equation}
\end{equation}
and take the derivative with respect to $\gamma_1$ \cite{jaynes2003}; this transforms Eq.~\eqref{eq:functional_equation} into $p(\gamma_1 \theta_1)+\gamma_1\,\theta_1\, \Dot{p}(\gamma_1 \theta_1) = 0$, where $\Dot{p}(z) \coloneqq d p(z)/dz$. In turn, this leads to the ordinary differential equation 
\begin{equation}
    p (z) + z\,\Dot{p}(z) = 0
\end{equation}
whose solution can be found straightforwardly as 
\begin{equation}
    \frac{dp(z)}{p(z)} = -\frac{dz}{z} \implies p(z) = \frac{\mathrm{const.}}{z} \propto \frac{1}{z}.
    \label{eq:complete_ignorance_app}
\end{equation}
Choosing $z \mapsto \theta_1$ leads to Jeffreys's prior $p(\theta_1) \propto 1/\theta_1$ for the first parameter. 
Since the same reasoning applies to the other parameters, we recover the result in the main text, i.e., $p(\boldsymbol{\theta}) \propto \prod_i (1/\theta_i)$.

% References
\bibliography{refs2022sep}

\end{document}